\documentclass[10pt, a4paper]{article}

\usepackage{arxiv}         
\usepackage[utf8]{inputenc}

\usepackage[english]{babel}
\usepackage[T1]{fontenc}    
\usepackage{hyperref}       
\usepackage{url}            
\usepackage{booktabs}       
\usepackage{amsfonts}       
\usepackage{microtype}      
\usepackage{amsmath}
\usepackage{amssymb}        
\usepackage{amsthm}         
\usepackage{dsfont}         
\usepackage{float}
\usepackage{graphicx}
\usepackage{stmaryrd}       
\usepackage{tikz}

\usepackage{newlfont}
\usepackage{indentfirst}
\usepackage{xcolor}
\usepackage{mathtools}
\usepackage{multirow}
\usepackage{cancel}
\usepackage{subcaption}
\usepackage{booktabs}
\usepackage{enumitem}
\allowdisplaybreaks

\newtheorem{theorem}{Theorem}
\newtheorem{proposition}{Proposition}

\newtheorem{assumption}{Assumption}

\newcommand{\indep}{\mbox{$\perp\!\!\!\perp$}}
\newcommand{\cb}{ \begin{eqnarray} }
	\newcommand{\ce}{ \end{eqnarray} }

\newcommand{\var}{ \mbox{\rm Var} }
\newcommand{\expit}{ \mbox{\rm expit} }

\DeclareMathOperator*{\argmin}{arg\,min}

\usepackage{natbib}
\setcitestyle{authoryear}

\usepackage{pgfplots}
\usepgfplotslibrary{colormaps}

\usepackage{csquotes}
\usepackage{todonotes}
\usepackage{soul}           
\usepackage{marvosym}       
\usepackage{caption}
\usepackage{subcaption}

\usepackage{algorithm}
\usepackage{algpseudocode}

\usepackage{lipsum}

\title{Robust evaluation of treatment effects in longitudinal studies with truncation by death or other intercurrent events}

\author{Georgi Baklicharov$^1$, Kelly Van Lancker$^{1,2}$, and Stijn Vansteelandt$^1$\vspace{0.5cm}\\
$^1$Department of Mathematics, Computer Science and Statistics,\\
Ghent University, Ghent, Belgium\\
$^2$Department of Mathematics and Data Science,\\
Vrije Universiteit Brussel, Brussels, Belgium \vspace{0.5cm}\\
\href{mailto:georgi.baklicharov@ugent.be}{\texttt{georgi.baklicharov@ugent.be}}}
\date{March 11, 2026}

\begin{document}

\tikzstyle{every node}=[draw=black,thick,anchor=west]
\tikzstyle{selected}=[draw=red,fill=red!30]
\tikzstyle{optional}=[dashed,fill=gray!50]

\maketitle

\begin{abstract}
Intercurrent events, such as treatment switching, rescue medication, dropout, or truncation by death, frequently complicate intention-to-treat analyses in randomized clinical trials. Existing causal inference frameworks typically target hypothetical or principal stratum estimands (e.g., survivor average causal effects), which rely on unverifiable assumptions and can be sensitive to unmeasured confounders or positivity violations. We propose a novel approach that mitigates this sensitivity by using only information measured prior to the intercurrent event. Our key idea is to compare treated and untreated individuals, matched on baseline covariates, at the most recent time point before either experiences an intercurrent event. We call these contrasts Pairwise Last Observation Time (PLOT) estimands. PLOT estimands are identified in randomized trials without structural assumptions, even under severe positivity violations. Although PLOT-based tests may theoretically be susceptible to residual selection bias, we show this bias vanishes under standard conditions and remains negligible in extensive simulations. We develop asymptotically efficient, model-free tests and treatment effect estimators using data-adaptive nuisance parameter estimation. We evaluate performance via simulation and apply the method to re-analyze the DEVOTE trial, affected by truncation by death. PLOT offers a robust, data-driven alternative for evaluating treatment efficacy in the presence of complex intercurrent events.
\end{abstract}

\section{Introduction}\label{sec: intro}
Many randomized trials are affected by intercurrent events (ICEs), such as treatment switching, rescue medication, dropout, truncation by death, etc., that occur after treatment initiation but before outcome measurement. These events often complicate evidence-based decision-making. A beneficial treatment, for example, may appear harmful in an intention-to-treat (ITT) analysis when patients on the control arm deteriorate and are therefore more often switched to a successful rescue medication. ITT analyses become especially subtle when study outcomes are truncated due to death. Standard analyses, e.g., within survivors, are then prone to selection bias and may in particular suggest treatment effects on the study outcome, even when treatment only impacts survival \citep{rosenbaum1984consequences}.

Over the past decades, a rich literature has emerged to address ICEs in clinical trials, notably unified by the ICH E9(R1) addendum on estimands and sensitivity analysis \citep{ICH2019}. A significant portion of this work focuses on hypothetical estimands which quantify what the (ITT) effect would have been \textit{had a different study protocol been used} in which ICEs were prevented (e.g., rescue treatment or treatment switching were not an option) \citep[e.g.,][]{olarte2023hypothetical,morgan2025applying} or the ICE were not affected by treatment (e.g., had rescue treatment only been given to control patients who would have been rescued if assigned to the experimental treatment arm) \citep[e.g.,][]{michiels2021novel}. Alternatively, the framework of separable effects \citep[e.g.,][]{stensrud2023conditional} shifts the focus toward a mechanistic understanding, quantifying the effect of \textit{a modified formulation of the study drug that does not affect the ICE}. While these mechanistic insights are valuable for drug development, the explicit reliance on hypothetical formulations may diverge from the primary regulatory objective of evaluating the drug in its current form. Principal stratum (PS) estimands \citep{Frangakis2004principal,robins2007discussions} avoid ``what-if'' thinking regarding protocols or formulations. This makes them relevant for the challenging problem of truncation by death \citep{chaix2012commentary}  (which can alternatively be addressed in terms of conditional separable effects \citep{stensrud2023conditional}). However, the clinical interpretability of PS estimands is often limited by the fact that they target a selective, unidentified patient population. For example, the survivor average causal effect (SACE) \citep{Robins1986, hayden2005estimator, tchetgentchetgen2014, wang2017identification} is restricted to the subgroup who would survive regardless of treatment assignment, a group that cannot be identified at baseline.

While these estimand frameworks offer valuable pathways for addressing ICEs, their utility is often constrained by sensitivity to the structural assumptions required for identification. This is particularly acute when strict protocol guidelines, such as deterministic rules for initiating rescue medication, render ICE occurrence nearly certain under specific clinical conditions \citep{michiels2021novel, olarte2023hypothetical}, resulting in near-positivity violations.
Furthermore, ICEs are frequently driven by time-varying prognostic factors that also influence the primary outcome. While hypothetical estimand and separable effects frameworks can theoretically accommodate such factors, they are routinely ignored in practice. In the principal stratum setting, \cite{qu2020general} develop a general framework that accounts for time-varying covariates; nevertheless, circumventing strong and implausible identification assumptions remains a fundamental challenge \citep{vansteelandt2024chasing}. Collectively, these difficulties threaten the robustness of the resulting inferences, raising serious concerns in regulatory contexts.

In previous work \citep{michiels2025adjusting,bowden2025instrumental}, these concerns were addressed via instrumental variable (IV) analyses. Here, we instead target estimands that extrapolate less away from the observed data, yielding more robust analyses. Composite estimands represent one such approach, but introduce their own conceptual challenges. For instance, \cite{mallinckrodt2019estimands} and \cite{lu2025estimating} incorporate the ICE directly into the outcome definition by assigning an unfavorable value (e.g., zero) to the outcome after death. This approach is a natural fit for recurrent event analyses, which typically evaluate the expected number of events accumulated until death or study completion, and shares the logic of standard competing risk analyses and last observation carried forward (LOCF) approaches. Alternative composite strategies have been proposed for quality-of-life (QoL) outcomes, where treatment evaluation is informed by the joint distribution of QoL and survival \citep[e.g.,][]{martinussen2026joint}.
While-alive or while-on-treatment estimands \citep{schmidli2023estimands} offer another alternative by normalizing the cumulative outcome by time elapsed prior to death or another ICE. The resulting ratio is intended to adjust for the longer observation time that may arise in one treatment arm due to improved survival (or lower ICE rates). However, the causal interpretation of this ratio and the conditions required to identify it from observed data have not been fully formalized \citep{janvin2024causal} and, like composite estimands more broadly, it obscures whether the treatment affects the outcome, the occurrence of ICEs, or both.

In this paper, we propose a different direction in the context of longitudinal studies, which sidesteps these concerns by instead comparing treated and untreated individuals – matched on baseline covariates – at the last time point before either individual in the pair experiences an ICE (e.g., dies). Unlike LOCF, while-alive or while-on-treatment estimands, this Pairwise Last Observation Time (PLOT) approach ensures that comparisons between treated and untreated individuals are anchored to the same time point, preventing treatment effect estimates from being distorted by differential survival or unequal times to ICEs across treatment arms. While this strategy offers a more grounded comparison, the considered contrasts may nonetheless remain susceptible to a form of residual selection bias. We give insight into the nature of this bias, but find it to be small in extensive simulation studies, and show that it vanishes under the conditions typically assumed by existing approaches. We construct asymptotically efficient, model-free tests of the null hypothesis of no treatment effect, using data-adaptive estimation of nuisance parameters. With additional structural model assumptions, our framework moreover delivers well interpretable, data-adaptive estimators of treatment effect. Our framework accommodates arbitrary (uncensored) outcomes measured in longitudinal clinical trials and observational studies, and arbitrary intercurrent events (including dropout and truncation by death).

The remainder of the paper is organized as follows. We introduce and motivate the proposed PLOT estimands in Section \ref{sec: estimands}, with identification and hypothesis testing strategies in Section \ref{sec: identest}. Section \ref{sec: simulations} reports simulation results to evaluate the finite-sample performance of the proposed methods in comparison with existing approaches. We then discuss the assumptions under which valid hypothesis testing is possible in Section \ref{sec: assumptions} and relate our estimands to hypothetical estimands in Section \ref{sec: connection} to facilitate interpretation of the treatment effect. The methodology is applied to the DEVOTE trial, a cardiovascular outcomes trial in type 2 diabetes subjects which was affected by truncation by death, in Section \ref{sec: devote}. Section \ref{sec: discussion} concludes with a discussion of the main findings, limitations and directions for future work. Technical proofs and supplementary derivations are deferred to the Appendices.

\section{Estimand}\label{sec: estimands}
Suppose data are available for a random sample of subjects $i=1,...,n$ on a baseline covariate vector $L_i$, binary treatment $A_i$ and outcome $Y_i(t)$ (categorical or continuous), planned to be measured at fixed visit times $t=0, 1, \ldots, \tau$ (with 0 denoting the baseline visit). Throughout, we restrict attention to randomized treatments. The proposed framework and theoretical results in Section \ref{sec: estcond} and Appendix A, however, extend naturally to observational studies. Let $T_i\in\{0,1,\ldots,\tau\}$ be the last time subject $i$ is observed prior to dropout, death or any other ICE. We denote the observed data for individual $i$ by $O_i$ and assume mutual independence of $\{O_i:i=1,...,n\}$; we will often drop the subject identifier $i$ when considering a random subject. Let $Y^a(t)$ and $T^a$ denote the counterfactual outcome at time $t$ and counterfactual time to ICE that would have been observed under treatment $a=0,1$, respectively. Note that $Y^a(t)$ is not well-defined when a subject dies prior to time $t$. Importantly, the estimands introduced in this section are so constructed that they never depend on such undefined potential outcomes.

The aforementioned LOCF estimand contrasts $E\left\{Y^1(\min(T^1,t)) - Y^{0}(\min(T^{0},t))\right\}$, which may differ from zero even when treatment does not affect the outcome process, since $T^1$ may differ from $T^0$. We therefore propose to contrast outcomes of treated and untreated individuals at the same time, namely the last time both remained ICE-free. In particular, consider drawing two random, independent individuals and assigning treatment to one (for whom we observe $(T^1, Y^1(t), L)$) but not to the other (for whom we observe $(T^{*0}, Y^{*0}(t), L^*)$). Then we will consider the estimands
\begin{equation}\label{PLOTestimand}
    \Psi_t = E\left\{Y^1(\min(T^1,T^{*0},t)) - Y^{*0}(\min(T^1,T^{*0},t))\right\}
\end{equation}
and
\begin{equation}\label{PLOTratio}
\frac{E\left\{Y^1(\min(T^1, T^{*0}, t))\right\}}{E\left\{Y^{*0}(\min(T^1, T^{*0}, t))\right\}}.
\end{equation}
We refer to these as Pairwise Last Observation Time (PLOT) estimands.
Alternatively, consider drawing 2 random, independent individuals with the same baseline covariates $L=L^*$, and assigning treatment to one but not to the other. Then we will consider
\begin{equation}\label{CPLOTestimand}
\Phi_t = E\left[ E\left\{Y^1(\min(T^1, T^{*0}, t)) - Y^{*0}(\min(T^1, T^{*0}, t))|L=L^*\right\}\right]
\end{equation}
or
\begin{equation}\label{CPLOTratio}
\frac{E\left[E \left\{Y^1(\min(T^1, T^{*0}, t))|L=L^*\right\}\right]}{E\left[E \left\{Y^{*0}(\min(T^1, T^{*0}, t))|L=L^*\right\}\right]}.
\end{equation} 
We refer to these as Conditional Pairwise Last Observation Time (CPLOT) estimands. They generally differ from the PLOT estimands, and are thus not collapsible, because individuals with the same covariates tend to have similar times to ICE, thereby resulting in later times at which both are ICE-free. CPLOT estimands thus tend to contrast outcomes at later time points than PLOT estimands, except when $L=\emptyset$ or $L\indep T$, in which case they coincide. We generally view them as preferable by ``truncating’' fewer measurements. 

By contrasting outcomes between pairs of treated and untreated individuals, (C)PLOT estimands bear some resemblance to the win ratio \citep{pocock2012winratio}, which assesses whether a randomly selected treated individual has a more favorable outcome than a randomly selected control. However, our use of pairing is not to rank outcomes, but to anchor the comparison of both individuals at a common, pre-ICE time point. In this sense, our estimands can be viewed as ``synchronous'' variants of the LOCF, competing risk and while-alive estimands, carrying forward the last outcome at which {\it both} individuals remained ICE-free. This facilitates a more formal separation of treatment effect mechanisms. However, whether these estimands equal zero when treatment has no effect on the outcome process is investigated in Section~\ref{sec: assumptions}.

\section{Identification and estimation}\label{sec: identest}

\subsection{PLOT estimand: simple estimator}\label{sec: estPLOT}
Under the randomization assumption that $A \indep (Y^a(t), T^a) $, for all $t$ and $ a = 0, 1 $, and consistency \citep{hernan2020causal}, we have that
\begin{align*}
E \left\{Y^1 (\min(T^1, T^{*0}, t))\right\} 
&= \sum_{s=0}^t E \left\{Y^1 (s)|\min(T^1,T^{*0},t)=s\right\} P\left\{\min(T^1,T^{*0},t)=s\right\} \\
&= \sum_{s=0}^t E \left\{Y(s)|A = 1, A^* = 0, \min(T, T^*, t) = s\right\} \\
&\quad \times P\left\{\min(T, T^*, t) = s|A = 1, A^* = 0\right\}.
\end{align*}

The unknown expectations and probabilities in this expression can be readily estimated by averaging across all pairs of treated and untreated individuals, as in
\begin{align}\label{MargUnadj1}
    \sum_{s=0}^t\frac{1}{\sum_{i,j} A_i(1-A_j)}\sum_{i,j} Y_i(s)A_i(1-A_j)I\left\{\min(T_i, T_j, t) = s\right\}.
\end{align}
Similarly, $E\left\{Y^{*0}(\min(T^1, T^{*0},t))\right\}$ can be estimated as (\ref{MargUnadj1}) upon replacing $Y_i(s)$ by $Y_j(s)$, so that an additive effect estimate is obtained as
\begin{align}\label{MargUnadjEffect}
    \sum_{s=0}^t\frac{1}{\sum_{i,j} A_i(1-A_j)}\sum_{i,j} \left\{Y_i(s)-Y_j(s)\right\}A_i(1-A_j)I\left\{\min(T_i, T_j, t) = s\right\}.
\end{align}
The standard error can be computed via the bootstrap. In Appendix A, we outline a semiparametric efficient estimation strategy that yields an efficient estimator incorporating baseline covariates and provides analytical standard errors for \eqref{MargUnadjEffect}.

\subsection{CPLOT estimand}\label{sec: estcond}
While the PLOT estimand can be readily estimated as (\ref{MargUnadjEffect}) by comparing all pairs of treated and untreated individuals, the curse of dimensionality prohibits a similar strategy for the CPLOT estimand: it is generally impossible to find treated and untreated individuals with exactly the same baseline covariates. This makes identification and estimation more challenging.
For $a=0,1$, denote $p_{a,s}(L)=P(T>s|A=a,L)$ for $s\in\{-1,0,\ldots,\tau\}$, and $\mu_{a,s,u}(L)=E\{Y(s)|A=a,L,T>u\}$ for $(s,u)\in\mathcal{I} = \{(s,u): s\in\{0,\ldots,\tau\},\ u\in\{s-1,s\}\}$.
\begin{proposition}\label{prop: CPLOTident}
    Under (conditional) randomization (or $ A \indep (Y^a(t), T^a)|L$ for $a=0,1$) and the consistency assumption, $E\left[ E\left\{Y^1(\min(T^1, T^{*0}, t))|L=L^*\right\}\right]$ can be identified as
\begin{align*}
    &\sum_{s=0}^t E\left\{p_{0,s-1}(L)p_{1,s-1}(L)\mu_{1,s,s-1}(L)\right\} - I(t>s)E\left\{p_{0,s}(L)p_{1,s}(L)\mu_{1,s,s}(L)\right\}.
\end{align*}
\end{proposition}
Based on this, we find the EIF of the CPLOT estimand $\Phi_t$ under the nonparametric model $\mathcal{P}$ as given by the following proposition.
\begin{proposition}\label{prop: CPLOTeif}
    The EIF of $\Phi_t$ under the nonparametric model $\mathcal{P}$ equals
\begin{align}\label{CPLOTestEIF}
    &\sum_{s=0}^t \zeta_{1,s,s-1}(O) -\zeta_{0,s,s-1}(O) - I(t>s)\left\{\zeta_{1,s,s}(O)-\zeta_{0,s,s}(O)\right\}-\Phi_t,
\end{align}
where
\begin{align*}
    &\zeta_{a,s,u}(O) = p_{1,u}(L)p_{0,u}(L)\mu_{a,s,u}(L) + \frac{I(A=1-a)}{P(A=1-a|L)}\left\{I(T>u) - p_{1-a,u}(L)\right\}p_{a,u}(L)\mu_{a,s,u}(L)\\
    &\qquad + p_{1-a,u}(L)\frac{I(A=a)}{P(A=a|L)}\left\{Y(s)I(T>u) - p_{a,u}(L)\mu_{a,s,u}(L)\right\},
\end{align*}
i.e., $\zeta_{a,s,u}(O)$ is the EIF of $E\left\{p_{1,u}(L)p_{0,u}(L)\mu_{a,s,u}(L)\right\}$ under $\mathcal{P}$.
\end{proposition}
Proofs are given in Appendix B. Under the semiparametric model $\mathcal{P}_\pi$ with a single restriction: $P(A=1|L)=P(A=1)\equiv \pi$, the EIF of $\Phi_t$ equals (see Appendix B)
\begin{align}\label{CPLOTestEIFsp}
    &\sum_{s=0}^t \zeta^*_{1,s,s-1}(O)-\zeta^*_{0,s,s-1}(O) - I(t>s)\left\{\zeta^*_{1,s,s}(O)-\zeta^*_{0,s,s}(O)\right\}-\Phi_t,
\end{align}
where $\zeta^*_{a,s,u}(O)$ is defined as $\zeta_{a,s,u}(O)$, but with $P(A=a|L)$ replaced by $\pi$. A consistent estimator for $\Phi_t$ is readily obtained as the value that makes the sample average of the influence functions zero:
\begin{equation}\label{CondDMLsp}
    \hat{\Phi}^*_t = \frac{1}{n}\sum_{i=1}^n\sum_{s=0}^t\hat{\zeta}^*_{1,s,s-1}(O_i) - \hat{\zeta}^*_{0,s,s-1}(O_i) - I(t>s)\left\{\hat{\zeta}^*_i(1,s,s)-\hat{\zeta}^*_{0,s,s}(O_i)\right\}.
\end{equation}
When using a $n^{1/2}$-consistent estimator of the propensity score, this estimator has the following asymptotic properties (proof given in Appendix B).
\begin{theorem}\label{th: CPLOT}
    $\sqrt{n}\left(\hat{\Phi}^*_t-\Phi_t\right)$ converges to a mean zero normal distribution with variance equal to the variance of (\ref{CPLOTestEIFsp}), when the following conditions are satisfied:
    \begin{enumerate}
        \item The propensity score is consistently estimated at $\sqrt{n}$-rate (e.g., as a sample proportion when $A\indep L$) and all other nuisance parameters are estimated on a sample independent to the one on which $\hat{\Phi}^*_\tau$ is evaluated (e.g., using cross-fitting),
        \item $p_{a,s}(L)\mu_{1,s,s}(L)$, $p_{a,s-1}(L)\mu_{1,s,s-1}(L)$ and $p_{a,s}(L)$ are $O_p(1)$ for $a=0,1$ and $s=0,1,\ldots,\tau$,
        \item the following terms are $o_p(1)$ for $a=0,1$:
        \begin{align*}
            &\sum_{s=0}^t\mathbb{P}\left[\left\{p_{a,s}(L)-\hat{p}_{a,s}(L)\right\}^2\right]^{1/2},\\
            &\sum_{s=0}^t\mathbb{P}\left[\{p_{a,s}(L)\mu_{a,s,s}(L)-\hat{p}_{a,s}(L)\hat{\mu}_{a,s,s}(L)\}^2\right]^{1/2},\\
            &\sum_{s=0}^t\mathbb{P}\left[\{p_{a,s-1}(L)\mu_{a,s,s-1}(L)-\hat{p}_{a,s-1}(L)\hat{\mu}_{a,s,s-1}(L)\}^2\right]^{1/2},
        \end{align*}
        \item the following terms are $o_p(n^{-1/2})$ for $a=0,1$:
        \begin{align*}
            &\sum_{s=0}^t\mathbb{P}\left[\{p_{1-a,s}(L)-\hat{p}_{1-a,s}(L)\}^2 \right]^{1/2}\mathbb{P}\left[ \{p_{a,s}(L)\mu_{a,s,s}(L)-\hat{p}_{a,s}(L)\hat{\mu}_{a,s,s}(L)\}^2\right]^{1/2},\\
            &\sum_{s=0}^t\mathbb{P}\left[\{p_{1-a,s-1}(L)-\hat{p}_{1-a,s-1}(L)\}^2 \right]^{1/2}\mathbb{P}\left[ \{p_{a,s-1}(L)\mu_{a,s,s-1}(L)-\hat{p}_{a,s-1}(L)\hat{\mu}_{a,s,s-1}(L)\}^2\right]^{1/2}.
        \end{align*}
    \end{enumerate}
\end{theorem}

This theorem suggests that the proposed estimator allows for using flexible, data-adaptive estimation strategies for the nuisance parameters $P(T>u|A=a,L)$ and $E\{Y(s)|A=a,L,T>u\}$, including variable selection, splines, and other forms of regularization, as well as modern machine learning methods or hybrid approaches that combine statistical and algorithmic learning. While such data-adaptive techniques usually induce non-negligible regularization bias and may deliver estimators with slow convergence rates \citep{van_der_laan_targeted_2011,chernozhukov_double/debiased_2018}, this does not affect the asymptotic behavior of $\hat{\Phi}^*_t$, which is root-$n$ consistent and asymptotically linear as long as the conditions in Theorem \ref{th: CPLOT} are satisfied.

A $(1-\alpha)100\%$ Wald confidence interval for $\Phi_t$ can thus be constructed as $\hat{\Phi}^*_t\pm z_{1-\alpha/2}\hat{v}^{1/2}$, with $z_{1-\alpha/2}$ the $(1-\alpha/2)$-quantile of the standard normal distribution and $\hat{v}$ given by 1 over $n$ times the sample variance of the EIF of ${\Phi}^*_t$, with all unknowns substituted by consistent estimators. A generalization of this result to general propensity score estimators is given in Appendix B.

A drawback relative to the estimator \eqref{MargUnadjEffect} is the requirement of rate conditions on the nuisance estimators (which are moreover stronger than those imposed on the semiparametric efficient estimator for $\Psi_t$ in Appendix A).
These can be achieved, for instance, when both estimators of the nuisance functions in the product converge at a rate faster than $n^{-1/4}$, or when one is estimated at parametric rate while the other is merely consistent. This compensation in convergence rates — often called rate double robustness — is, however, slightly more restrictive here than in the case of AIPW estimators of the ATE. 

To estimate the nuisance functions $p_{a,u}(L_i)$ and $\mu_{a,s,u}(L_i)$, we employ cross-fitting with two complementary strategies tailored to survival and longitudinal outcomes. First, conditional survival probabilities $p_{a,u}(L) = P(T > u| A=a, L)$ are estimated using survival random forests \citep{ishwaran2008random}, fitted within training folds and used to obtain out-of-fold predictions under treatment and control regimes ($a=1,0$). Second, for the conditional mean outcome regressions $\mu_{a,s,s}(L)=E\{Y(s)|A=a, T > s, L\}$ and $\mu_{a,s,s-1}(L)=E\{Y(s)|A=a, T > s-1, L\}$, we utilize Super Learner \citep{van2007super}, applied to appropriately constructed long-format data sets indexed by time $s$. Specifically, at each fold we stack observations across time points, restricting to those at risk at the relevant horizon, and fit ensemble regressions separately for the treated and control arms. In our implementation, the Super Learner library consists of main effect generalized linear models (with and without elastic-net regularization), random forests, and generalized additive models. Out-of-fold predictions are then generated at each $s \leq t$. This cross-fitted procedure ensures valid nuisance estimates while mitigating overfitting.

\section{Simulation experiments}\label{sec: simulations}
To evaluate finite-sample performance under the null of no treatment effect on the outcome, we conducted simulation experiments with 1000 runs, $\tau=5$, $n=250$, and a randomized treatment $A$ with $P(A=1)=0.5$. Propensity scores were estimated as sample proportions, and the remaining nuisance parameters using Super Learner and survival random forests, as detailed in Section \ref{sec: estcond}. We compared our proposals, with and without 5-fold cross-fitting (CF), with a SACE estimator, an IPCW estimator, a LOCF estimator and a naive estimator which calculates the ATE among survivors without covariate adjustment. Tables report the mean estimate, Monte Carlo standard deviation, average estimated standard error and coverage of 95\% Wald confidence intervals. Additional information and results, including the distribution of $T$ across arms and an observational study setting, appear in Appendix C. In all settings, outcomes after the ICE are set to NA.

In the first setting we consider a 10-dimensional baseline covariate vector $L\sim N(0,\Sigma)$, where $\Sigma$ is an autoregressive correlation matrix with correlation parameter 0.5. The outcome is
$Y(t) = 0.5 + 0.2t + L_1+L_2/2+L_3/3+L_8+L_9/2+L_{10}/3+ 2U^2 + u +\epsilon_{t}$, where $U\sim N(0,1)$ is unmeasured baseline covariate, $\epsilon_{t}\sim N(0,1)$ and $u\sim N(0,1)$ is a random intercept, all independent of any other variables. The observed time to ICE is $T=\min(\lfloor T_{ICE}\rfloor,\tau)$, where $T_{ICE}$ follows a Weibull distribution with shape 1.5 and scale parameter equal to $\exp(2.5+0.8A-L_4/4-L_5/5-L_6/6+L_8/4+L_9/5+L_{10}/6+0.2(2A-1)U^2)$. 

The second setting mimics a realistic treatment switching scenario, where patients switch to rescue treatment whenever a marker $Z(t)$ exceeds a threshold, inducing time-varying confounding and near-positivity violations, thereby posing a substantial challenge for existing methods: SACE estimators struggle with time-varying confounding and IPCW estimators are highly sensitive to positivity violations. The outcome is $Y(t)\sim N(0.5+0.2t+U,1)$, the marker $Z(t)\sim N(0.7At + 0.2t + U,1)$, and $T = \argmin_t\{Z(t)> 2\}$, where $U\sim N(0,1)$ is an unmeasured baseline covariate. The vector of baseline covariates $L$ is $(Z(0),Y(0))$.

In the third setting we consider a binary outcome with $Y(0)=0$ and $P(Y(t)=1|A,L)=\expit(0.5t + 5L + u)$, for $t\geq 1$, where $L\sim N(0,1)$ is a measured baseline covariate and $u\sim N(0,1)$ is a random intercept. The time to ICE $T_{ICE}$ follows a Weibull distribution with shape 1.5 and scale parameter equal to $\exp(2.5+0.8A+L)$.

In the setting 4 we consider a count outcome and evaluate our proposal for estimands (\ref{PLOTratio}) and (\ref{CPLOTratio}). Estimation and inference is as explained in Section \ref{sec: estcond} (see Appendix D for more details). $L$ and $T_{ICE}$ are as in setting 3. The outcome evolves through incremental changes over time, where the conditional mean of the outcome increment at each $t$ is given by $\lambda_t = \exp(-0.5 + 0.1t - 0.3L + 0.2\log(1 + Y(t-1)))$. At each time point, the outcome increment is drawn from a negative binomial distribution with dispersion 1.2 and mean $\lambda_t$, and the cumulative outcome at $t$ is obtained by summing these increments up to $t$. Here we compare only against a naive survivors-based estimator and LOCF.

\begin{table}[h!]
\centering
\begin{tabular}{lcccc}
  \hline
& Estimate  & SD  & SE  & Cov \\ 
  \hline
PLOT & -0.016 (-0.033) & 0.464 (0.407) & 0.474  (0.473) & 94.9 \\ 
  PLOTadj & 0.284 (-0.033) & 9.972 (0.188) & 0.414 (0.125) & 76.4 \\ 
  PLOTadj-CF & -0.013 (-0.015) & 0.210 (0.189) & 0.211 (0.207) & 95.7 \\ 
  CPLOT & 0.443 (-0.003) & 14.130 (0.200) & 0.526 (0.123) & 74.5 \\ 
  CPLOT-CF & 0.004 (0.001) & 0.216 (0.197) & 0.204 (0.201) & 93.6 \\ 
  SACE & 0.028 (0.038) & 0.485 (0.434) & 1.18$\times 10^{11}$ (0.859) & 98.9 \\ 
  IPCW & 0.070 (0.064) & 0.232 (0.219) & 0.205 (0.204) & 89.8 \\ 
  Survivors & 0.107 (0.098) & 0.510 (0.470) & 0.520 (0.517) & 94.6 \\ 
  LOCF & 0.150 (0.139) & 0.474 (0.422) & 0.484 (0.483) & 93.9 \\ 
   \hline
\end{tabular}
\caption{Simulation results for setting 1. Estimate: mean (median) of estimates; SD: standard deviation of estimates (robust sd using \texttt{cov.mcd} function in R); SE: mean (median) of estimated standard errors; Cov: coverage of 95\% confidence intervals. PLOT denotes unadjusted estimator (\ref{MargUnadjEffect}) and PLOTadj(-CF) denotes the adjusted alternative (with 5-fold cross-fitting) from Appendix A. CPLOT(-CF) denotes estimator (\ref{CondDMLsp}) (with 5-fold cross-fitting). SACE is the estimator proposed by \cite{hayden2005estimator}. IPCW is an inverse probability of censoring weighting estimator. ``Survivors'' is the treatment effect among individuals reaching the end of the study. LOCF is the last observation carried forward estimator.}
\label{tab: hd}
\end{table}

Table \ref{tab: hd} presents the simulation results for the first setting. All PLOT estimators are approximately unbiased with 95\% confidence intervals having coverage near the nominal level, except when no cross-fitting is used (which also leads to high variability due to some outliers). In contrast, IPCW, LOCF and the survivor-based estimator show moderate bias and high variance. IPCW is slightly biased and standard errors are underestimated, leading to undercoverage. The SACE estimator appears unbiased, but has very conservative standard error estimates, limiting its utility.

\begin{table}[h!]
\centering
\resizebox{\textwidth}{!}{%
\begin{tabular}{lcccccccccccc}
\hline
 & \multicolumn{4}{c}{Setting 2} & \multicolumn{4}{c}{Setting 3} & \multicolumn{4}{c}{Setting 4}\\
 & Est. & SD & SE & Cov & Est. & SD & SE & Cov & Est. & SD & SE & Cov \\
\hline
PLOT  & -0.001 & 0.152 & 0.156 & 95.3 & -0.008 & 0.056 & 0.054 & 94.2 &  0.063 & 0.112 & 0.111 & 90.9 \\
PLOTadj    & -0.003 & 0.123 & 0.100 & 88.5 & -0.008 & 0.049 & 0.033 & 80.2 &  0.064 & 0.111 & 0.084 & 79.9 \\
PLOTadj-CF & -0.004 & 0.122 & 0.126 & 95.8 & -0.010 & 0.050 & 0.049 & 94.4 &  0.064 & 0.111 & 0.110 & 90.0 \\
CPLOT      &  0.000 & 0.123 & 0.093 & 85.3 &  0.002 & 0.049 & 0.033 & 80.8 &  0.009 & 0.110 & 0.084 & 86.2 \\
CPLOT-CF   & -0.002 & 0.123 & 0.117 & 94.3 &  0.002 & 0.050 & 0.050 & 94.6 &  0.006 & 0.112 & 0.112 & 94.6 \\
SACE       & -0.280 & 0.305 & 0.316 & 84.3 &  0.007 & 0.062 & 0.061 & 94.3 & -&-&-&- \\
IPCW       & -0.382 & 0.371 & 0.222 & 54.5 &  0.002 & 0.070 & 0.060 & 91.8 & -&-&-&- \\
Survivors  & -0.691 & 0.286 & 0.291 & 36.6 & -0.053 & 0.068 & 0.066 & 85.3 &  0.079 & 0.119 & 0.118 & 88.8 \\
LOCF       & -0.223 & 0.161 & 0.165 & 72.9 &  0.027 & 0.062 & 0.062 & 92.3 &  0.241 & 0.122 & 0.120 & 48.5 \\
\hline
\end{tabular}%
}
\caption{Simulation results for settings 2 (treatment switching), 3 (binary outcome), and 4 (count outcome). Est.: mean of estimates; SD: standard deviation of estimates; SE: mean of estimated standard errors; Cov: coverage of 95\% confidence intervals. PLOT denotes unadjusted estimator (\ref{MargUnadjEffect}) and PLOTadj(-CF) denotes the adjusted alternative (with 5-fold cross-fitting) from Appendix A. CPLOT(-CF) denotes estimator (\ref{CondDMLsp}) (with 5-fold cross-fitting). SACE is the estimator proposed by \cite{hayden2005estimator}. IPCW is an inverse probability of censoring weighting estimator. ``Survivors'' is the treatment effect among individuals reaching the end of the study. LOCF is the last observation carried forward estimator. Dashes indicate methods not applied to Setting 4.}
\label{tab:rescuebinarycount}
\end{table}

Table \ref{tab:rescuebinarycount} presents the simulation results for settings 2, 3, and 4. Across both simulation settings, the PLOT estimators (with cross-fitting) show negligible bias and nominal coverage. This is so despite the PLOT estimand differing from zero in setting 3 and 4 due to the nonlinear link function (see Section~\ref{sec: assumptions}). Without cross-fitting, standard errors are underestimated, leading to reduced coverage. The adjusted PLOT and CPLOT estimators have the smallest variability among all methods. The gains are particularly evident in setting 2, where alternatives such as SACE, IPCW, LOCF and the survivor-based estimator suffer from bias and poor coverage due to violations of their assumptions. In setting 3, SACE and IPCW perform reasonably well in terms of bias, but have higher variance, and IPCW deviates from the nominal coverage level. The LOCF and survivor-based estimators are clearly biased. The results for setting 4 (count outcome) are very similar to setting 3. The CPLOT estimator with cross-fitting again exhibits the best performance. The PLOT estimators (both adjusted and unadjusted) deviate further from zero due to the nonlinear link function, but nonetheless show a huge improvement relative to LOCF by comparing individuals at the same time point.

\section{Assumptions for valid treatment effect testing} \label{sec: assumptions}
While we have argued that the (C)PLOT estimands facilitate a more formal separation of the underlying mechanisms of treatment effect than is possible with LOCF and while-alive estimands, they may deviate from zero even in the absence of a treatment effect on the outcome. This can be seen upon rewriting the PLOT estimand (\ref{PLOTestimand}) equivalently as $E\left\{Y^1(\min(T^1,T^{*0},t)) - Y^{0}(\min(T^{*1},T^{0},t))\right\}$, which compares potential outcomes of each individual under treatment and control at time points $\min(T^1,T^{*0},t)$ and $\min(T^{*1},T^{0},t)$ that are the same in distribution, though not necessarily identical. 

To illustrate, consider a simple data-generating model with no treatment effect on the outcome:
\begin{equation}\label{example}
    Y(t)=\alpha(t)+\beta_0 L+\beta_1 Lt+\gamma_0 U + \gamma_1 Ut + \epsilon_t,
\end{equation}
where $L$ and $U$ are baseline covariates and $\epsilon_t$ is an independent mean-zero error. Under this model, PLOT estimand $\Psi_t$ reduces to (see Appendix E) $\beta_1 E\left\{(L-L^*)\min(T^1,T^{*0},t)\right\} + \gamma_1 E\left\{(U-U^*)\min(T^1,T^{*0},t)\right\}$. It is zero when either treatment does not affect the time to ICE, $L$ and $U$ are independent of the time to ICE, or $\beta_1=\gamma_1=0$, i.e., when the additive effect of time on the outcome is not modified by any baseline covariates that also influence $T$. The CPLOT estimand $\Phi_t$ (conditioning on only $L$) reduces to $\gamma_1 E\left[E\left\{(U-U^*)\min(T^1,T^{*0},t)|L=L^*\right\}\right]$. It is zero under the weaker conditions that either treatment does not affect the time to ICE, $U\indep T|L$, or $\gamma_1=0$, i.e., when all such effect modifiers are measured at baseline. 

Table \ref{tab: sec5} summarizes simulation results (for 1000 runs) under model (\ref{example}) with $\epsilon_t\sim N(0,1)$, $T=\min(\lfloor T_{ICE}\rfloor,\tau)$ and $T_{ICE}$ Weibull with shape 1.5 and scale parameter equal to $\exp(2.5+0.8A+L+0.2(2A-1)U)$. $L$ and $U$ are generated from independent standard normal distributions, $\beta_0=1$, $\tau=5$, $n=250$ and $(\beta_1,\gamma_0,\gamma_1)\in\{(0,0,0),(0,1,0),(1,1,0),(1,1,1)\}$. The results confirm our previous findings: unmeasured common causes of outcome and $T$ are not problematic unless they modify the additive time effect on the outcome. When all such time effect modifiers are measured at baseline (i.e., if $(\beta_1,\gamma_0,\gamma_1)=(1,1,0)$), the CPLOT estimator averages to zero, while the PLOT estimator deviates. Likewise, problems can be expected in nonlinear outcome models, as was seen in simulation setting 3. Furthermore, if $(\beta_1,\gamma_0,\gamma_1)=(1,1,1)$, all methods fail, though the deviation remains small for CPLOT. Additional results in the Appendix C confirm that the CPLOT estimator is least sensitive to such violations, followed by the SACE estimator, which is typically much more variable.

\begin{table}[htbp]
\centering
\begin{tabular}{lcccccccc}
\hline
Method & Estimate & SD & SE & Cov & Estimate & SD & SE & Cov \\
\hline
& \multicolumn{4}{c}{$(\beta_1,\gamma_0,\gamma_1)=(0,0,0)$} & 
\multicolumn{4}{c}{$(\beta_1,\gamma_0,\gamma_1)=(0,1,0)$} \\
PLOT     & 0.003  & 0.159 & 0.161 & 96.1& -0.002 & 0.237 & 0.240 & 94.8 \\
PLOTadj-CF    & 0.009  & 0.212 & 0.108 & 95.4 & 0.002  & 0.143 & 0.141 & 96.2 \\
CPLOT-CF       & 0.009  & 0.220 & 0.106 & 94.3  & 0.016  & 0.144 & 0.139 & 93.9 \\
SACE        & 0.012  & 0.187 & 0.212 & 97.3  & 0.079  & 0.259 & 0.320 & 96.9 \\
IPCW        & -0.003 & 0.173 & 0.148 & 91.7   & 0.131  & 0.222 & 0.190 & 85.8 \\
Survivors   & -0.299 & 0.190 & 0.196 & 68.1  & -0.065 & 0.266 & 0.276 & 95.3 \\
LOCF        & 0.178  & 0.193 & 0.195 & 85.6     & 0.173  & 0.259 & 0.262 & 90.1 \\[2mm]
&\multicolumn{4}{c}{$(\beta_1,\gamma_0,\gamma_1)=(1,1,0)$} & 
\multicolumn{4}{c}{$(\beta_1,\gamma_0,\gamma_1)=(1,1,1)$} \\
PLOT     & -0.661 & 0.555 & 0.580 & 80.1      & -0.156 & 0.944 & 0.961 & 94.9 \\
PLOTadj-CF    & -0.670 & 0.225 & 0.229 & 15.4     & -0.141 & 0.532 & 0.540 & 93.9 \\
CPLOT-CF       & -0.004 & 0.183 & 0.178 & 94.6      & 0.144  & 0.491 & 0.481 & 94.1 \\
SACE        & 0.119  & 0.694 & 0.829 & 97.0    & 0.448  & 1.273 & 1.661 & 97.1 \\
IPCW        & 0.131  & 0.222 & 0.190 & 85.8    & 0.797  & 0.810 & 0.724 & 78.1 \\
Survivors   & -1.570 & 0.775 & 0.806 & 51.0   & -0.398 & 1.368 & 1.396 & 94.8 \\
LOCF        & -0.443 & 0.655 & 0.683 & 92.6    & 1.007  & 1.156 & 1.175 & 87.3 \\
\hline
\end{tabular}
\caption{Simulation results under the example model with $\beta_0=1$ and different values for $(\beta_1,\gamma_0,\gamma_1)$. Estimate: mean of estimates; SD: standard deviation of estimates; SE: mean of estimated standard errors; Cov: coverage of 95\% confidence intervals. PLOT denotes unadjusted estimator (\ref{MargUnadjEffect}) and PLOTadj(-CF) denotes the adjusted alternative (with 5-fold cross-fitting) from Appendix A. CPLOT(-CF) denotes estimator (\ref{CondDMLsp}) (with 5-fold cross-fitting). SACE is the estimator proposed by \cite{hayden2005estimator}. IPCW is an inverse probability of censoring weighting estimator. ``Survivors'' is the treatment effect among individuals reaching the end of the study. LOCF is the last observation carried forward estimator.}
\label{tab: sec5}
\end{table}

To formally examine the conditions under which the proposed estimands preserve the null hypothesis of no treatment effect, we define $Y^a(t)$ (with a slight abuse of notation) to be the outcome at time $t$ under treatment assignment $a$ in the hypothetical situation that no ICEs occur. The corresponding null hypothesis is $Y^1(t)=Y^0(t),\forall t$. Although such hypothetical reasoning may feel less natural in the presence of truncation by death, and moreover the proposed estimands do not rely on outcome measurements obtained after ICEs, it remains a common starting point for formulating assumptions, even in the truncation-by-death literature \citep{hayden2005estimator,ding2011identifiability}. We can now write
\begin{align*}
    \Psi_t&= \iiint yf_{Y^1(\min(v,v^*,t))|T^1}(y|v)f_{T^1}(v)f_{T^{*0}}(v^*)\,\mathrm{d}y\mathrm{d}v\mathrm{d}v^*\\
    &\qquad - \iiint yf_{Y^0(\min(v,v^*,t))|T^0}(y|v)f_{T^0}(v)f_{T^{*1}}(v^*)\,\mathrm{d}y\mathrm{d}v\mathrm{d}v^*\\
    &= \iiint yf_{Y^1(\min(v,v^*,t))}(y)f_{T^1}(v)c
    _1(F_{Y^1(\min(v,v^*,t))}(y),F_{T^1}(v))f_{T^0}(v^*)\,\mathrm{d}y\mathrm{d}v\mathrm{d}v^*\\
    &\qquad - \iiint yf_{Y^0(\min(v,v^*,t))}(y)f_{T^0}(v)c_0(F_{Y^0(\min(v,v^*,t))}(y),F_{T^0}(v))f_{T^1}(v^*)\,\mathrm{d}y\mathrm{d}v\mathrm{d}v^*,
\end{align*}
where $c_a(.,.)$ denotes the copula density of $(Y^a(t),T^a)$. Thus, $\Psi_t=0$ under the null whenever the dependence between $Y^1(t)$ and $T^1$ is the same as between $Y^0(t)$ and $T^0$ in the sense that, for all $t$, the copulas of $(Y^1(t),T^1)$ and $(Y^0(t),T^0)$ are equal. An analogous condition involving conditional copulas given $L$ suffices for the CPLOT estimand.
While we deem these assumptions to be relatively weak, especially when conditioning on $L$, they are nonetheless difficult to judge. 
Appendix E therefore shows that the CPLOT estimand is also guaranteed to be zero under the null hypothesis whenever at least one of the following assumptions holds.
\begin{assumption}\label{ass1}
    At least one of the following conditions is satisfied:
    \begin{enumerate}[label=(\alph*)]
    \item \label{ass1:T1isT0} $T^1 = T^0$ almost surely;
    \item \label{ass1:ipw} $\forall t:\, \underline{Y^a(t)}\indep (T^a>t)|L$, for $a=0,1$, where $\underline{Y^a(t)}$ is the vector consisting of $Y^a(t)$ and all future values of $Y^a(s),s>t$;
    \item \label{ass1:yt-y0} $\forall t:\, Y^a(t)-Y(0)\indep T^a|L$, for $a=0,1$.
\end{enumerate}
\end{assumption}

Condition \ref{ass1:T1isT0} states that the treatment does not affect ICE times. If satisfied, both estimands \eqref{PLOTestimand} and \eqref{CPLOTestimand} are trivially zero under the null. Condition \ref{ass1:ipw} is essentially the same as that invoked by IPW estimators of hypothetical estimands (which, unlike our proposal, invoke an additional positivity assumption). Condition \ref{ass1:yt-y0}, which applies to additive contrasts only (and is thus not relevant for estimand \eqref{CPLOTratio}), is satisfied in the absence of unmeasured common causes of outcome and $T$ that modify the additive time effect on the outcome. For the PLOT estimands to return zero under the null, the same assumption with $L=\emptyset$ suffices. The fact that this requirement is considerably stronger, motivates our preference for CPLOT estimands, despite their stronger rate conditions.

When hypothetical reasoning ``in the absence of ICEs'' is artificial, we may instead define the null as $Y^1(t)=Y^0(t),\forall t\leq\min(T^1, T^0)$. An alternative (principal stratum inspired) choice for the null is given in Appendix E. Both estimands continue to return zero under this null when $T^1=T^0$ almost surely. This continues to be the case under the following assumption (see the proof in the Appendix E):
\begin{assumption}\label{ass: sace}
    $\forall t:\, \{Y^a(t),Y^a(t+1)\}\indep T^{1-a}|T^a> t, L$, for $a=0,1$. 
\end{assumption} 
Assumption \ref{ass: sace} is essentially an explainable nonrandom survival assumption, as used in the PS literature \citep{hayden2005estimator}, though it does not suffice to identify the PS estimand. For instance, \cite{hayden2005estimator} additionally requires $(T^1\geq t)\indep (T^0\geq t)|L$, which is implausible \citep{vansteelandt2024chasing}. 

Finally, we emphasize that all assumptions outlined above are sufficient but not necessary. It follows that our testing proposal is valid under weaker assumptions than those imposed by competing methods.

\section{Estimation of hypothetical estimands}\label{sec: connection}

While CPLOT was specifically engineered to yield  robust hypothesis tests, interpreting its magnitude at a given time $t$ can be challenging. It typically acts as a conservative proxy for the hypothetical treatment effect that would be observed in the absence of ICEs. This is because CPLOT attenuates the treatment effect at time $t$ by incorporating contrasts from earlier time points for pairs experiencing prior ICEs.
To address this, the current section demonstrates how techniques inspired by IV methods can be employed to correct for this attenuation. Our focus here is on the CPLOT estimand on the additive scale; a corresponding derivation for the ratio-based estimand \eqref{CPLOTratio} is provided in Appendix F.

\subsection{Additive Treatment Effects}\label{sec: addlininterpret}
To build intuition, consider a scenario where the hypothetical treatment effect on the outcome grows linearly over time with a constant rate of effect accumulation $\psi$:
\begin{equation}\label{interpretLinear}
E(Y^1(s) | L) = E(Y^0(s) | L) + \psi s, \quad \forall s \in \{0, 1, \ldots, \tau\}
\end{equation}
Note that this is inherently a hypothetical assumption as it defines a contrast existing only in a counterfactual world where ICEs can be prevented. Under either of the conditions of Section \ref{sec: assumptions}, the CPLOT estimand \eqref{CPLOTestimand} then attenuates the treatment effect $\psi t$ to $\psi E\left[ E\left\{ \min(T^1, T^{*0}, t) | L=L^* \right\} \right]$. The dilution factor is the expected time until the first ICE occurs in a matched pair. The hypothetical effect $\psi$ can thus be recovered via the normalized estimand:
\begin{equation}\label{normalized_psi}
\frac{E\left[ E\left\{ Y^1(\min(T^1, T^{*0}, t)) - Y^{*0}(\min(T^1, T^{*0}, t)) | L=L^* \right\} \right]}{E\left[ E\left\{ \min(T^1, T^{*0}, t) | L=L^* \right\} \right]},
\end{equation}
which can be estimated using DML following the same principles used for estimating \eqref{CPLOTestimand}.

Moreover, by replacing \eqref{interpretLinear} with $E(Y^1(s) - Y^0(s)|T^1\geq s,T^0\geq s, L) = \psi s,\; \forall s \in \{0, 1, \ldots, \tau\}$, we can apply the exact same reasoning as above if we instead rely on assumption \ref{ass: sace}. In this case, $\psi t$ corresponds to the SACE at time $t$ and the CPLOT estimand can be interpreted as a diluted version of this effect.

\subsection{General Time-Varying Effects}
The linearity assumption in \eqref{interpretLinear} is restrictive, but can be relaxed by allowing for an arbitrary, time-specific hypothetical effect $\psi_s$:
\begin{equation}\label{interpretGeneral}
E(Y^1(s) | L) = E(Y^0(s) | L) + \psi_s, \quad \forall s \in \{0, 1, \ldots, \tau\}
\end{equation}
where $\psi_0 = 0$. Under the conditions of Section \ref{sec: assumptions}, the CPLOT estimand \eqref{CPLOTestimand} at time $t$ becomes a weighted average of these hypothetical effects: 
\[
E\left[ E\left\{ \psi_{\min(T^1, T^{*0}, t)} | L=L^* \right\} \right] = \sum_{s=0}^t\psi_sE\left[ P\left\{ \min(T^1, T^{*0}, t)=s | L=L^* \right\} \right].
\]
The vector $\psi = (\psi_1, \ldots, \psi_\tau) \in \mathbb{R}^{\tau}$ of hypothetical effects can thus be recovered by relating it to the CPLOT estimands via the following system of equations:
\begin{align*}
    & \Phi_t \equiv E\left[ E\left\{ Y^1(M_t) - Y^{*0}(M_t) | L=L^* \right\} \right] = E\left[E\left\{\psi_{M_t}|L=L^*\right\}\right], \quad \forall t \in \{1, \ldots, \tau\},
\end{align*}
where $M_t = \min(T^1, T^{*0}, t)$. In Appendix F, we find the solution to be
\begin{equation*}
\psi_t = \sum_{s=1}^t\frac{\Phi_s - \Phi_{s-1}}{E\left[P\left\{\min(T^1,T^{*0},s)=s|L=L^*\right\}\right]}.
\end{equation*}
Detailed estimation and inference procedures for $\psi$ are provided in Appendix F.

Replacing \eqref{interpretGeneral} with $E(Y^1(s) - Y^0(s)|T^1\geq s,T^0\geq s, L) = \psi_s,\; \forall s \in \{0, 1, \ldots, \tau\}$ allows us to apply the exact same reasoning as above by instead relying on assumption \ref{ass: sace}, such that $\Phi_t$ can be interpreted as a weighted average of survival average causal effects. Thus, depending on the assumptions one is willing to make, the CPLOT estimand can be interpreted as either a diluted hypothetical effect or a diluted SACE.

\section{Application to the DEVOTE trial}\label{sec: devote}
To illustrate the proposed methodology, we re-analyze data from the DEVOTE trial \citep{marso2017efficacy}, a randomized trial comparing the cardiovascular safety of insulin degludec (IDeg OD) and insulin glargine (IGlar OD) in patients with type 2 diabetes at high cardiovascular risk. While the  primary endpoint concerned major adverse cardiovascular events, our analysis targets a critical secondary outcome: the cumulative frequency of severe hypoglycemia, defined as episodes requiring third-party assistance.

The analysis was restricted to the same dataset as in \citep{marso2017efficacy} after removing subjects with missing data in the baseline covariates. This reduced the sample from 7,637 to 6,527 individuals because of missingness in baseline covariates. Notably, missingness was solely a byproduct of the anonymization process applied to the protected version of the dataset used for this analysis, and can thus safely be assumed to be missing completely at random. We define $Y(t)$ as the total number of severe hypoglycemic events experienced up to day $t$. The observation time $T_i$ for each subject was defined as the time from enrollment to the last planned follow-up visit (LPLV) for subjects who attended this visit, or, for subjects who did not complete follow-up, as the time to the last observed event among: last direct contact, last EAC-confirmed MACE prior to LPLV, and death prior to LPLV. For computational reasons, daily cumulative counts were converted to weekly cumulative outcomes $W(k) = Y(7k)$, and the last observed time was transformed to weeks as $\lfloor T_i/7 \rfloor$. Due to the trial's staggered enrollment, observation lengths varied. Only 5\% of the cohort remained under observation beyond 132 weeks. We therefore restricted our evaluation of the CPLOT estimand to $t \leq 132$. 

Baseline characteristics included age, BMI, estimated glomerular filtration rate, duration of diabetes, sex, insulin treatment, triglycerides, total cholesterol and HbA1c. Within the filtered cohort, 3,289 patients were assigned to IDeg OD ($A=1$), and 3,238 patients were assigned to IGlar OD ($A=0$). Mortality was low in both arms (171 and 192 deaths, respectively), and counterfactual survival curves are given in Appendix G. The counterfactual probability of being death-free at week 132 was estimated at $93.6\%$ (95\% CI: $92.6\%, 94.6\%$) for IDeg OD and $92.5\%$ (95\% CI: $91.3\%, 93.7\%$) for IGlar OD. The average cumulative event count per person (measured at the last available time point) was 0.071 in the IDeg OD arm and 0.116 in the IGlar OD arm.

We estimated CPLOT estimand \eqref{CPLOTestimand} to compare the mean number of severe hypoglycemic events while accounting for truncation by death and dropout. Propensity scores were estimated as sample proportions, while the remaining nuisance parameters were estimated using 3-fold cross-fitting, as detailed in Section \ref{sec: estcond}, but with GAMs instead of Super Learner for the estimation of the conditional mean outcome regressions $\mu_{a,s,s}(L)=E\{Y(s)|A=a, T > s, L\}$ and $\mu_{a,s,s-1}(L)=E\{Y(s)|A=a, T > s-1, L\}$.

\begin{figure}[htbp]
     \centering
     \begin{subfigure}[b]{0.8\linewidth}
         \centering
         \includegraphics[width=\textwidth]{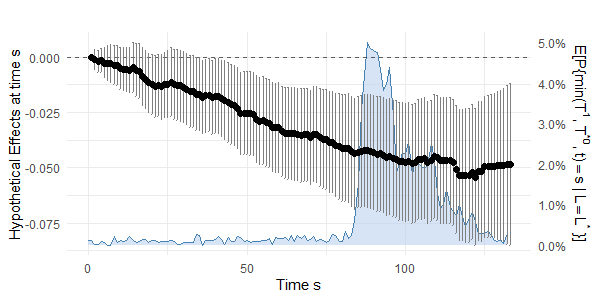}
         \caption{Additive Scale}
         \label{fig:devote_add}
     \end{subfigure}
     \hfill
     \begin{subfigure}[b]{0.8\linewidth}
         \centering
         \includegraphics[width=\textwidth]{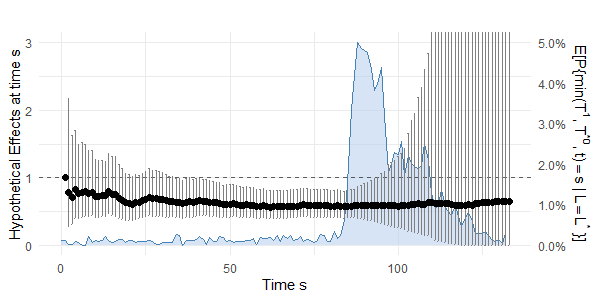}
         \caption{Multiplicative Scale}
         \label{fig:devote_mult}
     \end{subfigure}
        \caption{\textbf{Pointwise Treatment Contrasts $\psi_s$ and Estimated Density of First ICE Time in a Matched Pair.} Left axis: estimated contrasts $\psi_s$ (black points) with 95\% Wald CIs (grey bars), admitting a hypothetical (Assumption~\ref{ass1}) or SACE (Assumption~\ref{ass: sace}) interpretation; see Section~\ref{sec: connection}. Right axis: estimated density of $E[P\{\min(T^1, T^{*0}, t) = s \mid L=L^*\}]$ (shaded). Dashed line: null effect. Panel (b) y-axis restricted to $[0,3]$; later CIs truncated. Time $s$ in weeks from baseline.}
        \label{fig:devote_combined}
\end{figure}

The CPLOT estimand at $t=132$ on the additive scale was estimated as $-0.043$ (95\% CI: $(-0.069, -0.018)$; $p = 0.0007$), indicating a significant reduction in average total number of severe hypoglycemic events for IDeg OD compared to IGlar OD. The pointwise contrasts $\psi_s$ recovered via the inversion procedure of Section~\ref{sec: connection} are displayed in Figure~\ref{fig:devote_add}. Under Assumption \ref{ass1}, these estimates carry a hypothetical interpretation: the absolute reduction in average total number of severe hypoglycemic events on IDeg OD versus IGlar OD that would have been observed at each week $s$ had ICEs been prevented for all subjects. Alternatively, under the principal stratification assumption (Assumption~\ref{ass: sace}), the same estimates admit a SACE interpretation, representing the treatment contrast among individuals who would have remained ICE-free under either arm through week $s$.

A progressive divergence in the additive contrast is evident over the study period; while initial treatment effects are small, a steady downward trend emerges, indicating a growing advantage for IDeg OD. Notably, the uncertainty around these estimates increases after $s = 110$. Overlaying the estimated density of the first ICE time --- specifically, estimates for $E[P\{\min(T^1, T^{*0}, t) = s \mid L=L^*\}]$ at each $s$ --- reveals that the highest data density occurs between weeks 80 and 110 (Figure \ref{fig:devote_add}, shaded region), suggesting that the CPLOT estimand is primarily driven by the treatment dynamics during this high-density window. This peak coincides with the interval where the hypothetical effect stabilizes near $-0.045$, indicating that for an individual remaining under observation throughout this interval, IDeg OD reduced the expected event count by approximately $0.045$ (corresponding to roughly 293 episodes across the trial cohort) relative to IGlar OD, under a hypothetical regime that eliminates ICEs.

The SACE estimated via the \cite{hayden2005estimator} estimator yielded an effect of similar magnitude $-0.049$, suggesting that IDeg OD reduced the average total number of severe hypoglycemic events by approximately 0.05 per person compared to IGlar OD, within the principal stratum of individuals who would reach the end of the study without experiencing death or other ICEs under either treatment assignment. However, the associated 95\% confidence interval of $(-4.60, 4.50)$ is extremely wide, rendering the estimate uninformative in practice. The SACE estimate at $t = 132$ derived from Figure~\ref{fig:devote_add} under Assumption~\ref{ass: sace} --- corresponding to the last plotted point --- yielded $-0.048$ with a substantially narrower 95\% CI of $(-0.085,\, -0.012)$, demonstrating the efficiency gains afforded by the proposed approach. That both estimators converge to nearly identical point estimates provides informal reassurance that the result is robust to the choice of identifying assumption, though the two rest on substantively different identification strategies.

For reference, two naïve comparators were also computed. First, we implemented a LOCF analysis, resulting in an additive contrast of $-0.045$ (95\% CI: $(-0.072,-0.019)$; $p = 0.0008$). Second, we considered a survivors-only analysis, comparing mean last available outcomes among patients alive at $t=132$. This resulted in an additive contrast of $-0.049$ (95\% CI: $(-0.077,-0.021)$; $p=0.0006$). Both yield point estimates similar in magnitude to the CPLOT estimand, but their validity requires that survival and treatment assignment be independent.

The results on the multiplicative scale (Figure \ref{fig:devote_mult}), obtained as explained in Appendices D and F, reinforce these findings. Early in the study, the pointwise rate ratio $\psi_s$ remains near the null but exhibits a steady downward trend as $s$ increases, eventually stabilizing at a value indicative of a protective effect. The CPLOT estimand \eqref{CPLOTratio} at time $t=132$ was estimated as $0.60$ ($p=0.0004$) with a 95\% Wald confidence interval of $(0.45, 0.79)$, again indicating a significant difference in treatment effect in favor of insulin degludec. These results are broadly consistent with standard aggregate modeling approaches, with remarkably comparable precision. A Negative Binomial regression yielded a RR of $0.62$ (95\% CI: $0.48, 0.80$; $p < 0.001$), while an Andersen-Gill semiparametric model provided a hazard ratio of $0.63$ (95\% CI: $0.49, 0.82$; $p < 0.001$). An inverse probability weighted estimator targeting the hypothetical population-level effect at $t=132$ on the multiplicative scale yielded a rate ratio of 0.43 (95\% CI: $(-0.512,1.39)$; $p = 0.37075$, based on robust sandwich standard errors), which, while directionally consistent with the other approaches, suggests a somewhat stronger treatment effect under the assumptions implied by the IPW framework. However, unlike these global summary measures, the CPLOT approach reveals that the multiplicative estimates exhibit significantly wider confidence intervals at later time points ($s > 110$), reflecting the diminishing number of matched pairs at risk. Despite this late-stage variance, the point estimates during the period of highest event density provide evidence of a sustained relative benefit that is obscured by the single-number summaries of the NB and AG models.

\section{Discussion}\label{sec: discussion}

In this article, we proposed new methodology for analyzing longitudinal clinical trial data in the presence of ICEs. Causal analyses targeting hypothetical or principal stratum estimands are attractive in such settings, but their regulatory utility is limited by a lack of robustness, resulting from lack of data on time-varying confounders, lack of flexible methodology to handle them (for PS approaches), frequent (near-)positivity violations, and reliance on strong assumptions about the joint distribution of counterfactuals (for PS approaches). Motivated by this, we target a less ambitious estimand which avoids hypothetical reasoning in testing the null hypothesis of no treatment effect. The proposed (C)PLOT estimands compare outcomes of pairs of individuals at their last common ICE-free time, an idea that was briefly mentioned in the discussion of \cite{buyse2010generalized} on handling missing data in the estimation of the win ratio, avoiding extrapolation to unobserved post-ICE outcomes while targeting the full study population. 

While (C)PLOT estimands are identified under randomization and consistency alone, they make take non-zero values under the null due to selective timing of pairwise comparisons. We have demonstrated that this is not the case under the same assumptions as used by IPCW or PS approaches, and even under weaker conditions. To formalize sufficient conditions, we introduced a copula-based representation of the joint dependence between potential outcomes and time to ICE, which also provides a foundation for future sensitivity analyses. Initial sensitivity analyses (Appendix C) confirm that the CPLOT estimand exhibits considerably lower sensitivity to structural assumptions than competing approaches. This enhanced stability is a direct consequence of the CPLOT framework's avoidance of hypothetical reasoning; by staying closer to the observed data distribution rather than extrapolating to counterfactual scenarios, the estimand is more generally expected to reduce the influence of unverifiable structural assumptions.

Monte Carlo simulations showed good finite-sample performance, including in scenarios where competing approaches fail. When deviations from zero are expected, they are relatively small and typically smaller than those from alternative methods. Our estimators were also consistently more efficient than IPCW and PS estimators.

We have provided insight into the interpretation of the CPLOT estimand by removing effect attenuation. Future work will additionally explore alternative variants with enhanced interpretability, such as the probability that a treated individual has a more favorable outcome than an untreated individual in a random (matched) pair at their last common ICE-free time.

This pairwise perspective extends naturally to time-to-event settings with competing risks. Let $S$ denote the time to an event of interest and $T$ the time to a competing event. An analogous (C)PLOT strategy would contrast $E\left\{P\left(S^1 \leq \min(T^1, T^{*0}, \tau)|L=L^*\right)\right\}$ versus $E\left\{P\left(S^{*0} \leq \min(T^1, T^{*0}, \tau)|L=L^*\right)\right\}$, standardizing the ``window of opportunity'' for the event across the matched pair. Because matching on covariates typically creates a positive correlation between the survival times of the pair, this approach maximizes the follow-up window while disentangling the treatment effect on the event of interest from its effect on the competing risk.

Although the main focus has been on randomized treatments, the CPLOT methodology extends directly to observational studies when baseline covariates $L$ suffice to adjust for confounding of the effects of treatment on both the outcome and time to ICE. In such settings, a propensity overlap-weighted variant of the CPLOT estimand, i.e., \[E\left[ E\left\{Y^1(\min(T^1, T^{*0}, t)) - Y^{*0}(\min(T^1, T^{*0}, t))|L=L^*\right\}\frac{P(A=1|L)P(A=0|L)}{E\{P(A=1|L)P(A=0|L)\}}\right],\] may be preferable as it focuses on the subpopulation with the most balanced treatment assignment, mitigating the influence of extreme propensity scores.

\section*{Acknowledgements}

The authors are grateful to Marco Carone, discussions with whom gave inspiration for the proposed approach. They would also thank Linbo Wang for providing insight into the bias under the null, and to Jonathan Bartlett and Antoine Chambaz for useful discussions.

Finally, the authors express their gratitude to Jesper Madsen, Christian Bressen Pipper, Henrik Ravn, and Klaus Kähler Holst from Novo Nordisk for providing the data and for the many insightful discussions that helped shape this work.

\bibliographystyle{unsrtnat}
\bibliography{references}

\newpage 
{\Huge \textbf{Appendices}}
\section*{Appendix A: Efficient estimator of the PLOT estimand}
Using the same notation as in Section \ref{sec: identest}, we have the following identification result. 
\begin{proposition}\label{prop: PLOTident}
    Under (conditional) randomization ($A \indep (Y^a(t), T^a)|L$, for $a=0,1$) and the consistency assumption, $E\left\{Y^1 (\min(T^1, T^{*0}, t))\right\}$ can be identified as
\begin{equation}\label{AdjIdent}
    \sum_{s=0}^t E\{p_{0,s-1}(L)\}E\left\{p_{1,s-1}(L)\mu_{1,s,s-1}(L)\right\} - I(t>s)E\{p_{0,s}(L)\}E\left\{p_{1,s}(L)\mu_{1,s,s}(L)\right\}.
\end{equation}
\end{proposition}
\begin{proof}
    \begin{align*}
    &E \left\{Y^1 (\min(T^1, T^{*0}, t))\right\}= E\left[E\left\{Y^1 (\min(T^1, T^{*0}, t))|L,L^*\right\}\right] \\
    &= \sum_{s=0}^t E\left[E\left\{Y^1 (s)|\min(T^1,T^{*0},t)=s,L,L^*\right\} P\left\{\min(T^1,T^{*0},t)=s|L,L^*\right\}\right] \\
    &= \sum_{s=0}^t E\left(E\left[E \left\{Y^1(s)|\min(T^1, T^{*0}, t) = s,L,L^*\right\} P\left\{\min(T^1, T^{*0}, t) = s|L,L^*\right\}|L\right]\right),
\end{align*}
where
\begin{align*}
    &E\left[E \left\{Y^1(s)|\min(T^1, T^{*0}, t) = s,L,L^*\right\} P\left\{\min(T^1, T^{*0}, t) = s|L,L^*\right\}|L\right]\\
    &= \iint yf(Y^1(s) = y|L, T^1 \geq s)\frac{P(\min(T^1,T^{*0},t)=s|Y^1(s) = y, L,L^*=l^*, T^1 \geq s)}{P(\min(T^1,T^{*0},t)=s|L,L^*=l^*, T^1 \geq s)}\, dy\\
    &\qquad\times  P\left\{\min(T^1, T^{*0}, t) = s|L,L^*=l^*\right\} \, f_L(l^*)\,dl^*\\
    &= P(T^1\geq s|L)\iint yf(Y^1(s) = y|L, T^1 \geq s)\\
    &\qquad\times P(\min(T^1,T^{*0},t)=s|Y^1(s) = y, L,L^*=l^*, T^1 \geq s) \, dy\, f_L(l^*)\,dl^*\\
    &= P(T^1\geq s|L)\iint yf(Y^1(s) = y|L, T^1 \geq s) \big[
    P(T^0 > s - 1|L=l^*)\\
    &\quad- P(T^1 > s|Y^1(s) = y, L, T^1 \geq s) P(T^0 > s|L=l^*)I(t>s)
    \big] \, dy\, f_L(l^*)\,dl^*\\
    &= P(T^1\geq s|L)E\{P(T^0 > s - 1|L)\}\int yf(Y^1(s) = y|L, T^1 \geq s)\, dy\\
    &\qquad- P(T^1\geq s|L)E\{P(T^0 > s| L)\}I(t>s)\int yf(Y^1(s) = y|L, T^1 \geq s)\\
    &\qquad\qquad\times  P(T^1 > s|Y^1(s) = y, L, T^1 \geq s) \, dy\\
    &= P(T^1\geq s|,L)\Big[E\{P(T^0 > s - 1|L)\}E\{Y^1(s)|L, T^1 \geq s\}\\
    &\qquad- \int yf(Y^1(s) = y|L, T^1 \geq s)P(T^1 > s|Y^1(s) = y, L, T^1 \geq s) \, dy I(t>s)E\{P(T^0 > s|L)\}\Big]\\
    &= P(T^1> s-1|L)E\{P(T^0 > s - 1|L)\}E\{Y^1(s)|L, T^1> s-1\}\\
    &\qquad- \int yf(Y^1(s) = y|L, T^1 \geq s)P(T^1 > s|Y^1(s) = y, L, T^1 \geq s) \, dy \\
    &\qquad\qquad\times I(t>s)E\{P(T^0 > s|L)\}P(T^1\geq s|,L)\\
    &= P(T^1> s-1|L)E\{P(T^0 > s - 1|L)\}E\{Y^1(s)|L, T^1> s-1\}\\
    &\qquad- \int yf(Y^1(s) = y|L, T^1 \geq s,T^1 > s)P(T^1 > s|L, T^1 \geq s) \, dy \\
    &\qquad\qquad\times I(t>s)E\{P(T^0 > s|L)\}P(T^1\geq s|,L)\\
    &= P(T^1> s-1|L)E\{P(T^0 > s - 1|L)\}E\{Y^1(s)|L, T^1> s-1\}\\
    &\qquad- \int yf(Y^1(s) = y|L, T^1 \geq s,T^1 > s)\frac{P(T^1 > s|L)}{P(T^1 \geq s|L)} \, dy  I(t>s)E\{P(T^0 > s|L)\}P(T^1\geq s|,L)\\
    &= E\{P(T^0 > s - 1|L)\}P(T^1> s-1|L)E\{Y^1(s)|L, T^1> s-1\}\\
    &\qquad- I(t>s)E\{P(T^0>s|L)\}P(T^1> s|L)E\{Y^1(s)|L, T^1> s\}.
\end{align*}
Averaging over $L$ and summing over $s$ yields
\begin{align*}
    &\sum_{s=0}^tE\Big[E\{P(T^0 > s - 1|L)\}P(T^1> s-1|L)E\{Y^1(s)|L, T^1> s-1\}\\
    &\qquad- I(t>s)E\{P(T^0>s|L)\}P(T^1> s|L)E\{Y^1(s)|L, T^1> s\}\Big]\\
    &= \sum_{s=0}^tE\{P(T^0 > s - 1|L)\}E\left[P(T^1> s-1|L)E\{Y^1(s)|L, T^1> s-1\}\right]\\
    &\qquad- I(t>s)E\{P(T^0>s|L)\}E\left[P(T^1> s|L)E\{Y^1(s)|L, T^1> s\}\right]\\
    &= \sum_{s=0}^tE\{P(T > s - 1|A=0,L)\}E\left[P(T> s-1|A=1,L)E\{Y(s)|A=1,L, T> s-1\}\right]\\
    &\qquad- I(t>s)E\{P(T>s|A=0,L)\}E\left[P(T> s|A=1,L)E\{Y(s)|A=1,L, T> s\}\right].
\end{align*}
Similarly, we have that $E\left\{Y^{*0} (\min(T^1, T^{*0}, t))\right\}$ can be identified as
\begin{align*}
    & \sum_{s=0}^tE\{P(T > s - 1|A=1,L)\}E\left[P(T> s-1|A=0,L)E\{Y(s)|A=0,L, T> s-1\}\right]\\
    &\qquad- I(t>s)E\{P(T>s|A=1,L)\}E\left[P(T> s|A=0,L)E\{Y(s)|A=0,L, T> s\}\right].
\end{align*}
\end{proof}
Define $\eta_{a,s} = E\{p_{a,s}(L)\}$ and $\gamma_{a,s,u} = E\{p_{a,u}(L)\mu_{a,s,u}(L)\}$. Then, by the chain rule, the efficient influence function (EIF) of (\ref{AdjIdent}) under the nonparametric model $\mathcal{P}$ \citep{Pfanzagl1990,bickel1993efficient} equals
\begin{align}\label{PLOTestEIF}
    &\sum_{s=0}^t EIF_{\eta_{0,s-1}}\gamma_{1,s,s-1} + \eta_{0,s-1}EIF_{\gamma_{1,s,s-1}} - I(t>s)\left(EIF_{\eta_{0,s}}\gamma_{1,s,s} + \eta_{0,s}EIF_{\gamma_{1,s,s}}\right),
\end{align}
where $EIF_{\eta_{a,s}}$ and $EIF_{\gamma_{a,s,u}}$ are the EIFs of $\eta_{a,s}$ and $\gamma_{a,s,u}$, respectively, with
\begin{align*}
    EIF_{\eta_{a,s}} &= p_{a,s}(L) + \frac{I(A=a)}{P(A=a|L)}\left\{I(T>s) - p_{a,s}(L)\right\}-\eta_{a,s},\\
    EIF_{\gamma_{a,s,u}} &=  p_{a,u}(L)\mu_{a,s,u}(L)+ \frac{I(A=a)}{P(A=a|L)}\left\{Y(s)I(T>u) - p_{a,u}(L)\mu_{a,s,u}(L)\right\}-\gamma_{a,s,u}.
\end{align*}
Let the uncentered influence functions be $\varphi_{\eta_{a,s}}(O) = EIF_{\eta_{a,s}}+\eta_{a,s}$ and $\varphi_{\gamma_{a,s,u}}(O)=EIF_{\gamma_{a,s,u}}+\gamma_{a,s,u}$, where $O$ is the observed data vector of one individual.
Then the following proposition gives the EIF of $\Psi_t$ under $\mathcal{P}$.
\begin{proposition}\label{prop:eifum}
    The EIF of the PLOT estimand $\Psi_t$ under the nonparametric model $\mathcal{P}$ is
    \begin{equation*}
        \sum_{s=0}^t \xi_{s,s-1}(O) - I(t>s)\xi_{s,s}(O) - 2\Psi_t,
    \end{equation*}
    where
    \begin{equation*}
        \xi_{s,u}(O) = \varphi_{\eta_{0,u}}(O)\gamma_{1,s,u} + \eta_{0,u}\varphi_{\gamma_{1,s,u}}(O) - \varphi_{\eta_{1,u}}(O)\gamma_{0,s,u} -\eta_{1,u}\varphi_{\gamma_{0,s,u}}(O).
    \end{equation*}
\end{proposition}
It follows from Proposition \ref{prop:eifum} that a debiased machine learning \citep{chernozhukov_double/debiased_2018} estimator for $\Psi_t$ is given by
\begin{equation}\label{AdjDML}
    \hat{\Psi}_t = \frac{1}{2n}\sum_{i=1}^n\sum_{s=0}^t\hat{\xi}_{s,s-1}(O_i) - I(t>s)\hat{\xi}_{s,s}(O_i),
\end{equation}
where $\hat{\xi}_{s,u}(O)=\hat{\varphi}_{\eta_{0,u}}(O)\hat{\gamma}_{1,s,u} + \hat{\eta}_{0,u}\hat{\varphi}_{\gamma_{1,s,u}}(O) - \hat{\varphi}_{\eta_{1,u}}(O)\hat{\gamma}_{0,s,u} -\hat{\eta}_{1,u}\hat{\varphi}_{\gamma_{0,s,u}}(O)$. Here, $\hat{\eta}_{a,s}$ and $\hat{\gamma}_{a,s,u}$ are debiased machine learning (AIPW) estimators for $\eta_{a,s}$ and $\gamma_{a,s,u}$ respectively, obtained as the sample averages of 
\begin{align*}
    \hat{\varphi}_{\eta_{a,s}}(O_i) &= \hat{p}_{a,s}(L_i) + \frac{I(A_i=a)}{\hat{P}(A_i=a|L_i)}\left\{I(T_i>s) - \hat{p}_{a,s}(L_i)\right\},\\
    \hat{\varphi}_{\gamma_{a,s,u}}(O_i) &= \hat{p}_{a,u}(L_i)\hat{\mu}_{a,s,u}(L_i)+ \frac{I(A_i=a)}{\hat{P}(A_i=a|L_i)}\left\{Y_i(s)I(T_i>u) - \hat{p}_{a,u}(L_i)\hat{\mu}_{a,s,u}(L_i)\right\},
\end{align*}
where all nuisance parameters are substituted by consistent estimators, indicated with a hat. 
The variance of this estimator can be consistently estimated as 1 over $n$ times the sample variance of $\sum_{s=0}^t\hat{\xi}_{s,s-1}(O_i) - I(t>s)\hat{\xi}_{s,s}(O_i)$.

Consider now the semiparametric model $\mathcal{P}_\pi$, which assumes that $A\indep L$, as guaranteed under simple randomization. Then $P(A=a|L)=P(A=a)$, $\eta_{a,s}=E\{P(T>s|A=a,L)\}=P(T>s|A=a)$ and $\gamma_{a,s,u} = E[E\{Y(s)I(T>u)|A=a,L\}] = E\{Y(s)I(T>u)|A=a\}$. Let $\varphi^*_{\eta_{a,s}}$ and $\varphi^*_{\gamma_{a,s,u}}$ denote the uncentered influence functions of $\eta_{a,s}$ and $\gamma_{a,s,u}$, respectively, where we replace $P(A=a|L)$ by $P(A=a)$. Define $\xi^*_{s,u}$ as $\xi_{s,u}$ but with $\varphi^*_{\eta_{a,s}}$ and $\varphi^*_{\gamma_{a,s,u}}$ in lieu of $\varphi_{\eta_{a,s}}$ and $\varphi_{\gamma_{a,s,u}}$. The EIF of $\Psi_t$ under $\mathcal{P}_\pi$ is given in the following proposition.
\begin{proposition}\label{prop: PLOTspEIF}
    The EIF of the PLOT estimand $\Psi_t$ under the semiparametric model $\mathcal{P}_\pi$ with a single restriction: $P(A=1|L)=P(A=1)\equiv \pi$ is
    \begin{align}\label{PLOTestEIFsp}
        &\sum_{s=0}^t \xi^*_{s,s-1}(O) - I(t>s)\xi^*_{s,s}(O) - 2\Psi_t,
    \end{align}
    which is readily obtained from (\ref{PLOTestEIF}) upon substituting $P(A=a|L)$ by $P(A=a)$.
\end{proposition}
\begin{proof}
    Simplifying the EIF of $\Psi_t$ (under $\mathcal{P}$) according to $A\indep L$, yields the EIF under the semiparametric model $\mathcal{P}_\pi$. Indeed, for an arbitrary influence function $\varphi$, the set of influence functions under $\mathcal{P}_\pi$ is given by (see e.g., \cite{tsiatis2006semiparametric})
    \[
        \{\varphi + (A-\pi)h(L): h(L)\in L_2(L)\},
    \]
    where $L_2(L)$ denotes the set of al square integrable functions of $L$. The EIF is the one with minimum variance and is given by $\varphi + (A-\pi)h^*(L)$, with 
    \[
        h^*(L) = \frac{E\{(A-\pi)\varphi|L\}}{E\{(A-\pi)^2\}}.
    \]
    Now, upon choosing $\varphi$ equal to the EIF under $\mathcal{P}$, which can be further simplified according to $A\indep L$, makes it is easy to verify that $h^*(L)=0$. Consequently, the chosen influence function $\varphi$ was already the efficient influence function under this semiparametric model.
\end{proof}
Consequently, the estimator 
\begin{equation}\label{AdjDMLsp}
    \hat{\Psi}^*_t = \frac{1}{2n}\sum_{i=1}^n\sum_{s=0}^t\hat{\xi}^*_{s,s-1}(O_i) - I(t>s)\hat{\xi}^*_{s,s}(O_i),
\end{equation}
is asymptotically efficient within this restricted semiparametric model, and thus in particular at least as efficient as \eqref{AdjDML} when both $P(T>u|A=a,L)$ and $E\{Y(s)|A=a,L,T>u\}$ are consistently estimated. This estimator exhibits model double robustness in the sense that it is asymptotically unbiased when either $P(T>u|A=a,L)$ and $E\{Y(s)|A=a,L,T>u\}$, or the propensity score, are consistently estimated. As the latter can be guaranteed when $A\indep L$ (in which case the propensity score can be consistently estimated as a sample proportion), (\ref{AdjDMLsp}) is an asymptotically unbiased estimator for $\Psi_t$, regardless of whether $P(T>u|A=a,L)$ and $E\{Y(s)|A=a,L,T>u\}$ are consistently estimated.

To prove that $\hat{\Psi}^*_t$ is asymptotically unbiased if (i) models for either $P(T>u|A=a,L)$ and $E\{Y(s)|A=a,L,T>u\}$, or (ii) the propensity score, are correctly specified, we focus on one representative component of the estimator, noting that the argument extends directly to the full expression:
\begin{align*}
    E\left\{\hat{\xi^*}_{s,u}(O)\right\} &= E\left\{\hat{\varphi}^*_{\eta_{0,u}}(O)\hat{\gamma}_{1,s,u} + \hat{\eta}_{0,u}\hat{\varphi}^*_{\gamma_{1,s,u}}(O) - \hat{\varphi}^*_{\eta_{1,u}}(O)\hat{\gamma}_{0,s,u} -\hat{\eta}_{1,u}\hat{\varphi}^*_{\gamma_{0,s,u}}(O)\right\}\\
    &= E\left[\hat{p}_{0,s}(L) + \frac{1-A}{1-\hat{P}(A=1)}\left\{I(T>s) - \hat{p}_{0,s}(L)\right\}\right]\hat{\gamma}_{1,s,u}\\
    &\qquad + \hat{\eta}_{0,u}E\left[ \hat{p}_{1,u}(L)\hat{\mu}_{1,s,u}(L)+ \frac{A}{\hat{P}(A=1)}\left[Y(s)I(T>u) - \hat{p}_{1,u}(L)\hat{\mu}_{1,s,u}(L)\right] \right]\\
    &\qquad - E\left[\hat{p}_{1,s}(L) + \frac{A}{\hat{P}(A=1)}\left\{I(T>s) - \hat{p}_{1,s}(L)\right\}\right]\hat{\gamma}_{0,s,u}\\
    &\qquad - \hat{\eta}_{1,u}E\left[ \hat{p}_{0,u}(L)\hat{\mu}_{0,s,u}(L)+ \frac{1-A}{1-\hat{P}(A=1)}\left[Y(s)I(T>u) - \hat{p}_{0,u}(L)\hat{\mu}_{0,s,u}(L)\right] \right]\\
    &= E\left[\hat{p}_{0,s}(L) + \frac{1-P(A=1)}{1-\hat{P}(A=1)}\left\{p_{0,s}(L) - \hat{p}_{0,s}(L)\right\}\right]\hat{\gamma}_{1,s,u}\\
    &\qquad + \hat{\eta}_{0,u}E\left[ \hat{p}_{1,u}(L)\hat{\mu}_{1,s,u}(L)+ \frac{P(A=1)}{\hat{P}(A=1)}\left[p_{1,u}(L)\mu_{1,s,u}(L) - \hat{p}_{1,u}(L)\hat{\mu}_{1,s,u}(L)\right] \right]\\
    &\qquad - E\left[\hat{p}_{1,s}(L) + \frac{P(A=1)}{\hat{P}(A=1)}\left\{p_{1,s}(L) - \hat{p}_{1,s}(L)\right\}\right]\hat{\gamma}_{0,s,u}\\
    &\qquad - \hat{\eta}_{1,u}E\left[ \hat{p}_{0,u}(L)\hat{\mu}_{0,s,u}(L)+ \frac{1-P(A=1)}{1-\hat{P}(A=1)}\left[p_{0,u}(L)\mu_{0,s,u}(L) - \hat{p}_{0,u}(L)\hat{\mu}_{0,s,u}(L)\right] \right].
\end{align*}
It is readily seen that this expression simplifies to 
\begin{align*}
    & E\left[p_{0,s}(L)\}\right]\gamma_{1,s,u}+ \eta_{0,u}E\left[ p_{1,u}(L)\mu_{1,s,u}(L) \right]- E\left[p_{1,s}(L)\right]\gamma_{0,s,u}- \eta_{1,u}E\left[ p_{0,u}(L)\mu_{0,s,u}(L)  \right]\\
    &= 2\eta_{0,u}\gamma_{1,s,u} - 2\eta_{1,u}\gamma_{0,s,u}
\end{align*}
in the limit if models for either $p_{a,u}(L)$ and $\mu_{a,s,u}(L)$, or the propensity score $P(A=1)$, are correctly specified. This establishes that $\hat{\Psi}^*_t$ is asymptotically unbiased under either condition (i) or (ii).

The resulting estimation strategy is of interest, even when choosing $L=\emptyset$. In this case, $\hat{\Psi}_t$ reduces to 
\begin{align}\label{UnadjEst}
\begin{split}
    & \frac{1}{n}\sum_{i=1}^n\sum_{s=0}^tY_i(s)I(T_i>s-1)\left\{
    \frac{A_i}{\hat{P}(A=1)}\hat{p}_{0,s-1} -\frac{1-A_i}{\hat{P}(A=0)}\hat{p}_{1,s-1}\right\}\\
     &\qquad\qquad-I(t>s)Y_i(s)I(T_i>s)\left\{\frac{A_i}{\hat{P}(A=1)}\hat{p}_{0,s}-\frac{1-A_i}{\hat{P}(A=0)}\hat{p}_{1,s}\right\},
\end{split}
\end{align}
where $\hat{p}_{a,s}=\hat{P}(T>s|A=a)$.
This estimator is algebraically equivalent to (\ref{MargUnadjEffect}), but can be computed in $O(n)$ time rather than $O(n^2)$. Its variance can be estimated using the same influence function-based approach, but with $L=\emptyset$. 

To formalize the asymptotic properties of $\hat{\Psi}^*_t$, we introduce some additional notation. Let $\mathbb{P}_n$ denote the empirical measure (i.e., the sample average). For a function $f(O)$ of the data $O$, we write $\mathbb{P}\{f(O)\} = \int f(O)\mathbb{P}(O)dO$, where $\mathbb{P}(O)$ is the data density. Importantly, $\mathbb{P}\{\hat{f}(O)\}$ averages over $O$, rather than over the estimator $\hat{f}$. When the propensity score is consistently estimated at parametric rate (e.g., as a sample proportion), we have the following asymptotic result.
\begin{theorem}\label{th: PLOT}
    When $A\indep L$, $\sqrt{n}\left(\hat{\Psi}^*_t-\Psi_t\right)$ converges to a mean zero normal distribution with variance equal to the variance of (\ref{PLOTestEIFsp}), when the following conditions are satisfied:
    \begin{enumerate}
        \item The propensity score is consistently estimated at $n^{1/2}$-rate (e.g., as a sample proportion)
        and all other nuisance parameters are estimated on a sample independent of the one on which $\hat{\Psi}^*_\tau$ is evaluated (e.g., using cross-fitting), and
        \item the following convergence conditions hold for $a=0,1$:
        \begin{align*}
            &\sum_{s=0}^t\mathbb{P}\left\{p_{a,s}(L)-\hat{p}_{a,s}(L)\right\}=o_p(1),\\
            &\sum_{s=0}^t\mathbb{P}\{p_{a,s}(L)\mu_{a,s,s}(L)-\hat{p}_{a,s}(L)\hat{\mu}_{a,s,s}(L)\}=o_p(1),\\
            &\sum_{s=0}^t\mathbb{P}\{p_{a,s-1}(L)\mu_{a,s,s-1}(L)-\hat{p}_{a,s-1}(L)\hat{\mu}_{a,s,s-1}(L)\}=o_p(1).
        \end{align*}
    \end{enumerate}
\end{theorem}
\begin{proof}
    The proof is limited to a study of the second order remainder in the von Mises expansion. We refer to standard proofs as in \cite{chernozhukov_double/debiased_2018} and \cite{vansteelandt2022assumption} for a study of the empirical process term under cross-fitting.
    Define 
    \begin{align*}
        \lambda_{a,s,u} &= E\{P(T>u|A=1-a,L)\}E\left[P(T> u|A=a,L)E\{Y(s)|A=a,L,T>u\}\right]\\
        &= \eta_{1-a,u}\gamma_{a,s,u}.
    \end{align*}
    Using the notation from the main paper, we focus on the remainder for $\lambda_{1,s,s}$. From this we can easily generalize the results to $\hat{\Psi}^*_t$. 
    
    The second order remainder term in the von Mises expansion for $\lambda_{1,s,s}$ is given by \citep{hines2021demystifying}
    \begin{align*}
        &\hat{\lambda}_{1,s,s} - \lambda_{1,s,s} + \mathbb{P}\left[ \hat{p}_{0,s}(L) + \frac{1-A}{1-\hat{\pi}}\{I(T>s)-\hat{p}_{0,s}(L)\} - \hat{\eta}_{0,s}\right]\hat{\gamma}_{1,s,s}\\
        & + \hat{\eta}_{0,s}\mathbb{P}\left[ \hat{p}_{1,s}(L)\hat{\mu}_{1,s,s}(L) + \frac{A}{\hat{\pi}}\{Y(s)I(T>s)-\hat{p}_{1,s}(L)\hat{\mu}_{1,s,s}(L)\} - \hat{\gamma}_{1,s,s}\right]\\
        &= -\hat{\lambda}_{1,s,s} - \lambda_{1,s,s} + \mathbb{P}\left[ \hat{p}_{0,s}(L) + \frac{1-\pi}{1-\hat{\pi}}\{p_{0,s}(L)-\hat{p}_{0,s}(L)\}\right]\hat{\gamma}_{1,s,s}\\
        & + \hat{\eta}_{0,s}\mathbb{P}\left[ \hat{p}_{1,s}(L)\hat{\mu}_{1,s,s}(L) + \frac{\pi}{\hat{\pi}}\{p_{1,s}(L)\mu_{1,s,s}(L)-\hat{p}_{1,s}(L)\hat{\mu}_{1,s,s}(L)\}\right]\\
        &= -\hat{\eta}_{0,s}\hat{\gamma}_{1,s,s} - \eta_{0,s}\gamma_{1,s,s} + \mathbb{P}\left[ p_{0,s}(L) + \left(\frac{1-\pi}{1-\hat{\pi}}-1\right)\{p_{0,s}(L)-\hat{p}_{0,s}(L)\} \right]\hat{\gamma}_{1,s,s}\\
        & + \hat{\eta}_{0,s}\mathbb{P}\left[ p_{1,s}(L)\mu_{1,s,s}(L) + \left(\frac{\pi}{\hat{\pi}}-1\right)\{p_{1,s}(L)\mu_{1,s,s}(L)-\hat{p}_{1,s}(L)\hat{\mu}_{1,s,s}(L)\}\right]\\
        &= -\hat{\eta}_{0,s}\hat{\gamma}_{1,s,s} - \eta_{0,s}\gamma_{1,s,s} + \eta_{0,s}\hat{\gamma}_{1,s,s} + \mathbb{P}\left[\left(\frac{1-\pi}{1-\hat{\pi}}-1\right)\{p_{0,s}(L)-\hat{p}_{0,s}(L)\} \right]\hat{\gamma}_{1,s,s}\\
        & + \hat{\eta}_{0,s}\gamma_{1,s,s} + \hat{\eta}_{0,s}\mathbb{P}\left[ \left(\frac{\pi}{\hat{\pi}}-1\right)\{p_{1,s}(L)\mu_{1,s,s}(L)-\hat{p}_{1,s}(L)\hat{\mu}_{1,s,s}(L)\}\right]\\
        &= (\eta_{0,s}-\hat{\eta}_{0,s})(\hat{\gamma}_{1,s,s} - \gamma_{1,s,s}) + \mathbb{P}\left[\left(\frac{1-\pi}{1-\hat{\pi}}-1\right)\{p_{0,s}(L)-\hat{p}_{0,s}(L)\} \right](\hat{\gamma}_{1,s,s}-\gamma_{1,s,s})\\
        & + (\hat{\eta}_{0,s}-\eta_{0,s})\mathbb{P}\left[ \left(\frac{\pi}{\hat{\pi}}-1\right)\{p_{1,s}(L)\mu_{1,s,s}(L)-\hat{p}_{1,s}(L)\hat{\mu}_{1,s,s}(L)\}\right]\\
        &+\mathbb{P}\left[\left(\frac{1-\pi}{1-\hat{\pi}}-1\right)\{p_{0,s}(L)-\hat{p}_{0,s}(L)\} \right]\gamma_{1,s,s}\\
        &+ \eta_{0,s}\mathbb{P}\left[ \left(\frac{\pi}{\hat{\pi}}-1\right)\{p_{1,s}(L)\mu_{1,s,s}(L)-\hat{p}_{1,s}(L)\hat{\mu}_{1,s,s}(L)\}\right]\\
        &= (\eta_{0,s}-\hat{\eta}_{0,s})(\hat{\gamma}_{1,s,s} - \gamma_{1,s,s})  + \left(\frac{1-\pi}{1-\hat{\pi}}-1\right)\mathbb{P}\{p_{0,s}(L)-\hat{p}_{0,s}(L)\} (\hat{\gamma}_{1,s,s}-\gamma_{1,s,s})\\
        & + (\hat{\eta}_{0,s}-\eta_{0,s})\left(\frac{\pi}{\hat{\pi}}-1\right)\mathbb{P}\{p_{1,s}(L)\mu_{1,s,s}(L)-\hat{p}_{1,s}(L)\hat{\mu}_{1,s,s}(L)\}\\
        &+\left(\frac{1-\pi}{1-\hat{\pi}}-1\right)\mathbb{P}\{p_{0,s}(L)-\hat{p}_{0,s}(L)\} \gamma_{1,s,s}\\
        &+ \eta_{0,s}\left(\frac{\pi}{\hat{\pi}}-1\right)\mathbb{P}\{p_{1,s}(L)\mu_{1,s,s}(L)-\hat{p}_{1,s}(L)\hat{\mu}_{1,s,s}(L)\}.
    \end{align*}
    Only in the last step, we use that $A\indep L$ ($P(A=1|L)=\pi$). For the remainder of the proof, we will assume that $\hat{\pi}$ is a $n^{1/2}$-consistent estimator of $P(A=1)$ (e.g., the sample proportion), that $0<\hat{\pi}<1$, and that $\eta_{a,s}$ and $\gamma_{a,s,u}$ are bounded.
    
    $\hat{\eta}_{a,s}$ is $n^{1/2}$-consistent if $(\hat{\pi}-\pi)\mathbb{P}\left\{p_{a,s}(L)-\hat{p}_{a,s}(L)\right\}=o_p(n^{-1/2})$, i.e., if $\mathbb{P}\left\{p_{a,s}(L)-\hat{p}_{a,s}(L)\right\}=o_p(1)$. 
    Similarly, $\hat{\gamma}_{1,s,s}$ is $n^{1/2}$-consistent if $\mathbb{P}\{p_{a,s}(L)\mu_{a,s,s}(L)-\hat{p}_{a,s}(L)\hat{\mu}_{a,s,s}(L)\}=o_p(1)$.
    
    We need the remainder to be $o_p(n^{-1/2})$. The second and third term in the expansion above are lower order terms. For the fourth term we need that $\mathbb{P}\left\{p_{0,s}(L)-\hat{p}_{0,s}(L)\right\}=o_p(1)$. For the fifth term, we need that $\mathbb{P}\{p_{1,s}(L)\mu_{1,s,s}(L)-\hat{p}_{1,s}(L)\hat{\mu}_{1,s,s}(L)\}=o_p(1)$. These conditions imply that $\hat{\eta}_{0,s}$ and $\hat{\gamma}_{1,s,s}$ are $n^{1/2}$-consistent, such that first term in the expansion is $o_p(n^{-1/2})$.
    
    Consequently, we need the following terms to be $o_p(1)$.
    \begin{align*}
        &\sum_{s=0}^t\mathbb{P}\left\{p_{a,s}(L)-\hat{p}_{a,s}(L)\right\},\\
        &\sum_{s=0}^t\mathbb{P}\{p_{a,s}(L)\mu_{a,s,s}(L)-\hat{p}_{a,s}(L)\hat{\mu}_{a,s,s}(L)\},\\
        &\sum_{s=0}^t\mathbb{P}\{p_{a,s-1}(L)\mu_{a,s,s-1}(L)-\hat{p}_{a,s-1}(L)\hat{\mu}_{a,s,s-1}(L)\}.
    \end{align*}
    
    Under these rate assumptions, we have that 
    \[
    \sqrt{n}\left(\hat{\Psi}^*_t-\Psi_t\right) = \frac{1}{n}\sum_{i=1}^n\sum_{s=0}^t \xi^*_{s,s-1}(O_i) - I(t>s)\xi^*_{s,s}(O_i) - 2\Psi_t + o_p(1),
    \]
    which converges to a mean zero normal distribution with variance equal to the variance of (\ref{PLOTestEIFsp}) by the central limit theorem.
\end{proof}
This result extends naturally to observational studies, provided that baseline covariates $L$ suffice to adjust for confounding of the effects of treatment on both the outcome and time to ICE. The estimator remains essentially unchanged, but the rate requirements would be that all of the following terms are $o_p(n^{-1/2})$.
    \begin{align*}
        &\sum_{s=0}^t\mathbb{P}\left[\left\{\pi(L)-\hat{\pi}(L)\right\}^2 \right]^{1/2}\mathbb{P}\left[\left\{p_{a,s}(L)-\hat{p}_{a,s}(L)\right\}^2 \right]^{1/2},\\
        &\sum_{s=0}^t\mathbb{P}\left[\left\{\pi(L)-\hat{\pi}(L)\right\}^2 \right]^{1/2}\mathbb{P}\left[\{p_{a,s}(L)\mu_{a,s,s}(L)-\hat{p}_{a,s}(L)\hat{\mu}_{a,s,s}(L)\}^2\right],\\
        &\sum_{s=0}^t\mathbb{P}\left[\left\{\pi(L)-\hat{\pi}(L)\right\}^2 \right]^{1/2}\mathbb{P}\left[\{p_{a,s-1}(L)\mu_{a,s,s-1}(L)-\hat{p}_{a,s-1}(L)\hat{\mu}_{a,s,s-1}(L)\}^2\right],
    \end{align*}
where $\pi(L) = P(A=1|L)$.

This theorem suggests that the proposed estimator allows for using flexible, data-adaptive estimation strategies for the nuisance parameters $P(T>u|A=a,L)$ and $E\{Y(s)|A=a,L,T>u\}$, including variable selection, splines, and other forms of regularization, as well as modern machine learning methods or hybrid approaches that combine statistical and algorithmic learning. While such data-adaptive techniques usually induce non-negligible regularization bias and may deliver estimators with slow convergence rates \citep{van_der_laan_targeted_2011,chernozhukov_double/debiased_2018}, this does not affect the asymptotic behavior of $\hat{\Psi}^*_t$, which is root-$n$ consistent and asymptotically linear (a property likewise attained for AIPW estimators of the ATE in simple randomized trials).
We can thus construct a $(1-\alpha)100\%$ Wald confidence interval for $\Psi_t$ as $\hat{\Psi}^*_t\pm z_{1-\alpha/2}\hat{v}^{1/2}$, with $z_{1-\alpha/2}$ the $(1-\alpha/2)$-quantile of the standard normal distribution and $\hat{v}$ given by 1 over root-$n$ times the sample standard deviation of the EIF of ${\Psi}^*_t$, with all unknowns substituted by consistent estimators. 

Notably, when the estimator is based on the known rather than estimated randomization probability, condition 2 of Theorem \ref{th: PLOT} vanishes entirely. We nonetheless recommend estimating the propensity scores as this improves efficiency and reduces variance when other nuisance parameters are not consistently estimated  \citep{rotnitzky2010note}, and may more generally give finite-sample benefits. A drawback, however, is that the aforementioned confidence interval may fail to cover at the nominal rate when inconsistent nuisance parameter estimators are used. This can be remedied by additionally accounting for the uncertainty in estimating the propensity score (as in \cite{kelly2024automated}).

\newpage
\section*{Appendix B: Proof of propositions and theorems in Section \ref{sec: estcond}}
\subsection*{Proof of Proposition \ref{prop: CPLOTident}}
Under the randomization (or more generally, $ A \indep (Y^a(t), T^a)|L$) and consistency assumption, $E\left\{Y^1 (\min(T^1, T^{*0}, t))\right\}$ can be identified as (\ref{AdjIdent}). Indeed, the components of the CPLOT estimand, can be identified as follows:
\begin{align*}
    &E\left[E\left\{Y^1 (\min(T^1, T^{*0}, t))|L=L^*\right\}\right] \\
    &= \sum_{s=0}^t E\left[E\left\{Y^1 (s)|\min(T^1,T^{*0},t)=s,L=L^*\right\} P\left\{\min(T^1,T^{*0},t)=s|L=L^*\right\}\right] \\
    &=\sum_{s=0}^t \iint yf(Y^1(s) = y|\min(T^1,T^{*0},t)=s,L=l,L^*=l, T^1 \geq s)\,dy\\
    &\qquad\qquad \times P\left\{\min(T^1, T^{*0}, t) = s|L=l,L^*=l\right\} \, f_L(l)\,dl\\
    &=\sum_{s=0}^t \iint yf(Y^1(s) = y|L=l, T^1 \geq s)\frac{P(\min(T^1,T^{*0},t)=s|Y^1(s) = y, L=l,L^*=l, T^1 \geq s)}{P(\min(T^1,T^{*0},t)=s|L=l,L^*=l, T^1 \geq s)}\, dy\\
    &\qquad\qquad \times P\left\{\min(T^1, T^{*0}, t) = s|L=l,L^*=l\right\} \, f_L(l)\,dl\\
    &=\sum_{s=0}^t \iint yf(Y^1(s) = y|L=l, T^1 \geq s)P(\min(T^1,T^{*0},t)=s|Y^1(s) = y, L=l,L^*=l, T^1 \geq s)\, dy\\
    &\qquad\qquad \times P\left\{T^1\geq s|L=l\right\} \, f_L(l)\,dl\\
    &= \sum_{s=0}^t \iint yf(Y^1(s) = y|L=l, T^1 \geq s)\left\{ P(T^{*0}\geq s|L^*=l) \right.\\
    &\qquad\left. - I(t>s)P(T^1>s|Y^1(s) = y, L=l, T^1 \geq s)P(T^{*0}>s|L^*=l)\right\}\, dy P(T^1\geq s|L=l) \, f_L(l)\,dl\\
    &= \sum_{s=0}^t E\left[E\left\{Y^1(s)|T^1>s-1,L\right\}P(T^1>s-1|L)P(T^0>s-1|L)\right]\\
    &\qquad - I(t>s)\iint yf(Y^1(s) = y|L=l, T^1 \geq s, T^1>s)P(T^1>s|L=l, T^1 \geq s)\\
    &\qquad\qquad\times P(T^{*0}>s|L^*=l)\, dy P(T^1\geq s|L=l) \, f_L(l)\,dl\\
    &= \sum_{s=0}^t E\left[E\left\{Y^1(s)|T^1>s-1,L\right\}P(T^1>s-1|L)P(T^0>s-1|L)\right]\\
    &\qquad - I(t>s)\iint yf(Y^1(s) = y|L=l,T^1>s)\frac{P(T^1>s|L=l)}{P(T^1\geq s|L=l)}\\
    &\qquad\qquad\times P(T^{*0}>s|L^*=l)\, dy P(T^1\geq s|L=l) \, f_L(l)\,dl\\
    &= \sum_{s=0}^t E\left[E\left\{Y^1(s)|T^1>s-1,L\right\}P(T^1>s-1|L)P(T^0>s-1|L)\right]\\
    &\qquad - I(t>s)E\left[E\left\{Y^1(s)|T^1>s,L\right\}P(T^1>s|L)P(T^0>s|L)\right]\\
    &= \sum_{s=0}^t E\left[P(T>s-1|A=1,L)P(T>s-1|A=0,L)E\left\{Y(s)|A=1,T>s-1,L\right\}\right]\\
    &\qquad - I(t>s)E\left[P(T>s|A=1,L)P(T>s|A=0,L)E\left\{Y(s)|A=1,T>s,L\right\}\right].
\end{align*}
Similarly, $E\left[E\left\{Y^{*0} (\min(T^1, T^{*0}, t))|L=L^*\right\}\right]$ can be identified as
\begin{align*}
    &\sum_{s=0}^t E\left[P(T>s-1|A=1,L)P(T>s-1|A=0,L)E\left\{Y(s)|A=0,T>s-1,L\right\}\right]\\
    &\qquad - I(t>s)E\left[P(T>s|A=1,L)P(T>s|A=0,L)E\left\{Y(s)|A=0,T>s,L\right\}\right],
\end{align*}
which proves Proposition \ref{prop: CPLOTident}.

\subsection*{Proof of Proposition \ref{prop: CPLOTeif}}
Since $E\left[ E\left\{Y^1(\min(T^1, T^{*0}, t))|L=L^*\right\}\right]$ can be identified as
\begin{align*}
    &\sum_{s=0}^t E\left\{p_{1,s-1}(L)p_{0,s-1}(L)\mu_{1,s,s-1}(L)\right\} - I(t>s)E\left\{p_{1,s}(L)p_{0,s}(L)\mu_{1,s,s}(L)\right\},
\end{align*}
we first derive the efficient influence function of 
\begin{align*}
    \theta_{s,u} &= E\left[P(T>u|A=1,L)P(T> u|A=0,L)E\{Y(s)|A=1,L,T>u\}\right]\\
    &= E\left[P(T> u|A=0,L)E\{Y(s)I(T>u)|A=1,L\}\right]
\end{align*} 
under the nonparametric model $\mathcal{P}$. By the chain rule, we have that the pathwise derivative of $\theta_{s,u}$ is
\begin{align*}
    & P(T> u|A=0,L)E\{Y(s)I(T>u)|A=1,L\} - \theta_{s,u}\\
    & + \frac{1-A}{P(A=0|L)}\left\{I(T>s) - P(T>s|A=0,L) \right\}E\{Y(s)I(T>u)|A=1,L\}\\
    &+ P(T>s|A=0,L)\frac{A}{P(A=1|L)}\left[Y(s)I(T>s)-E\{Y(s)I(T>u)|A=1,L\}\right]\\
    &= P(T> u|A=0,L)P(T> u|A=1,L)E\{Y(s)|A=1,L,T>u\} - \theta_{s,u}\\
    & + \frac{1-A}{P(A=0|L)}\left\{I(T>s) - P(T>s|A=0,L) \right\}P(T> u|A=1,L)E\{Y(s)|A=1,L,T>u\}\\
    &+ P(T>s|A=0,L)\frac{A}{P(A=1|L)}\left[Y(s)I(T>s)-P(T> u|A=1,L)E\{Y(s)|A=1,L,T>u\}\right]
\end{align*}
Consequently, the EIF of 
\[\Phi_t = E\left[E\left\{Y^1 (\min(T^1, T^{*0}, t))|L=L^*\right\} - E \left\{Y^{*0} (\min(T^1, T^{*0}, t))|L=L^*\right\}\right]\]
under $\mathcal{P}$ is
\begin{align*}
    &\sum_{s=0}^t P(T>s-1|A=1,L)P(T>s-1|A=0,L)\\
    &\qquad\times\left[E\left\{Y(s)|A=1,T>s-1,L\right\}-E\left\{Y(s)|A=0,T>s-1,L\right\}\right]\\
    &\qquad + P(T>s-1|A=1,L)\frac{1-A}{P(A=0|L)}\left\{ I(T>s-1) - P(T>s-1|A=0,L) \right\}\\
    &\qquad\qquad\times E\left\{Y(s)|A=1,T>s-1,L\right\}\\
    &\qquad+ P(T>s-1|A=0,L)\frac{A}{P(A=1|L)}\left[Y(s)I(T>s-1)\right.\\
    &\qquad\qquad\left. -P(T>s-1|A=1,L)E\left\{Y(s)|A=1,T>s-1,L\right\}\right]\\
    &\qquad - P(T>s-1|A=0,L)\frac{A}{P(A=1|L)}\left\{ I(T>s-1) - P(T>s-1|A=1,L) \right\}\\
    &\qquad\qquad\times E\left\{Y(s)|A=0,T>s-1,L\right\}\\
    &\qquad-P(T>s-1|A=1,L)\frac{1-A}{P(A=0|L)} \left[Y(s)I(T>s-1)\right.\\
    &\qquad\qquad\left.- P(T>s-1|A=0,L) E\left\{Y(s)|A=0,T>s-1,L\right\}\right]\\
    & - I(t>s)\Bigg(P(T>s|A=1,L)P(T>s|A=0,L)\\
    &\qquad\times\left[E\left\{Y(s)|A=1,T>s,L\right\}-E\left\{Y(s)|A=0,T>s,L\right\}\right]\\
    &\qquad + P(T>s|A=1,L)\frac{1-A}{P(A=0|L)}\left\{ I(T>s) - P(T>s|A=0,L) \right\} E\left\{Y(s)|A=1,T>s,L\right\}\\
    &\qquad+ P(T>s|A=0,L)\frac{A}{P(A=1|L)}\left[Y(s)I(T>s) -P(T>s|A=1,L)E\left\{Y(s)|A=1,T>s,L\right\}\right]\\
    &\qquad - P(T>s|A=0,L)\frac{A}{P(A=1|L)}\left\{ I(T>s) - P(T>s|A=1,L) \right\} E\left\{Y(s)|A=0,T>s,L\right\}\\
    &\qquad-\frac{1-A}{P(A=0|L)}P(T>s|A=1,L) \left[Y(s)I(T>s)- P(T>s|A=0,L) E\left\{Y(s)|A=0,T>s,L\right\}\right]\Bigg)\\
    &- \Phi_t.
\end{align*}
This proves Proposition \ref{prop: CPLOTeif}.

Proving the claim that simplifying the EIF of $\Phi_t$ (under $\mathcal{P}$) according to $A\indep L$, yields the EIF under the semiparametric model $\mathcal{P}_\pi$, is completely analogous to the proof of Proposition \ref{prop: PLOTspEIF}.

\subsection*{Proof of Theorem \ref{th: CPLOT}}
The proof is limited to a study of the second order remainder in the von Mises expansion. We refer to standard proofs as in \cite{chernozhukov_double/debiased_2018} and \cite{vansteelandt2022assumption} for a study of the empirical process term under cross-fitting.

To derive the rate conditions for $\Phi_t$, we will focus on
\begin{align*}
    \theta_{s,s} &= E\left[P(T>s|A=0,L)P(T> s|A=1,L)E\{Y(s)|A=1,L,T>s\}\right]\\
    &= E\{p_{0,s}(L)p_{1,s}(L)\mu_{1,s,s}(L)\}.
\end{align*}
The second order remainder term in the von Mises expansion for $\theta_{s,s}$ is given by \citep{hines2021demystifying}
\begin{align*}
    &\hat{\theta}_{s,s} - \theta_{s,s} + \mathbb{P}\left[ \hat{p}_{0,s}(L)\hat{p}_{1,s}(L)\hat{\mu}_{1,s,s}(L) \right] + \mathbb{P}\left[ \hat{p}_{1,s}(L)\hat{\mu}_{1,s,s}(L)\frac{1-A}{1-\hat{\pi}}\{I(T>s)-\hat{p}_{0,s}(L)\} \right]\\
    & + \mathbb{P}\left[\hat{p}_{0,s}(L)\frac{A}{\hat{\pi}}\{Y(s)I(T>s)-\hat{p}_{1,s}(L)\hat{\mu}_{1,s,s}(L)\}\right] - \hat{\theta}_{s,s}\\
    &= \mathbb{P}\left[ \hat{p}_{0,s}(L)\hat{p}_{1,s}(L)\hat{\mu}_{1,s,s}(L) \right] + \mathbb{P}\left[ \hat{p}_{1,s}(L)\hat{\mu}_{1,s,s}(L)\frac{1-\pi}{1-\hat{\pi}}\{p_{0,s}(L)-\hat{p}_{0,s}(L)\} \right]\\
    & + \mathbb{P}\left[\hat{p}_{0,s}(L)\frac{\pi}{\hat{\pi}}\{p_{1,s}(L)\mu_{1,s,s}(L)-\hat{p}_{1,s}(L)\hat{\mu}_{1,s,s}(L)\}\right] - \theta_{s,s}\\
    &= \mathbb{P}\left[ \hat{p}_{1,s}(L)\hat{\mu}_{1,s,s}(L)\left(\frac{1-\pi}{1-\hat{\pi}}-1\right)\{p_{0,s}(L)-\hat{p}_{0,s}(L)\} \right]\\
    & + \mathbb{P}\left[\hat{p}_{0,s}(L)\left(\frac{\pi}{\hat{\pi}}-1\right)\{p_{1,s}(L)\mu_{1,s,s}(L)-\hat{p}_{1,s}(L)\hat{\mu}_{1,s,s}(L)\}\right]\\
    & +\mathbb{P}\left[ \hat{p}_{1,s}(L)\hat{\mu}_{1,s,s}(L)\{p_{0,s}(L)-\hat{p}_{0,s}(L)\} \right]\\
    & + \mathbb{P}\left[\hat{p}_{0,s}(L)\{p_{1,s}(L)\mu_{1,s,s}(L)-\hat{p}_{1,s}(L)\hat{\mu}_{1,s,s}(L)\}\right]\\
    &+\mathbb{P}\left[ \hat{p}_{0,s}(L)\hat{p}_{1,s}(L)\hat{\mu}_{1,s,s}(L) \right]- \mathbb{P}\{p_{0,s}(L)p_{1,s}(L)\mu_{1,s,s}(L)\}\\
    &= \mathbb{P}\left[ \hat{p}_{1,s}(L)\hat{\mu}_{1,s,s}(L)\left(\frac{1-\pi}{1-\hat{\pi}}-1\right)\{p_{0,s}(L)-\hat{p}_{0,s}(L)\} \right]\\
    & + \mathbb{P}\left[\hat{p}_{0,s}(L)\left(\frac{\pi}{\hat{\pi}}-1\right)\{p_{1,s}(L)\mu_{1,s,s}(L)-\hat{p}_{1,s}(L)\hat{\mu}_{1,s,s}(L)\}\right]\\
    & +\mathbb{P}\left[ \{\hat{p}_{1,s}(L)\hat{\mu}_{1,s,s}(L)-p_{1,s}(L)\mu_{1,s,s}(L)\}\{p_{0,s}(L)-\hat{p}_{0,s}(L)\} \right]\\
    &= \mathbb{P}\left[ \left\{\hat{p}_{1,s}(L)\hat{\mu}_{1,s,s}(L)-p_{1,s}(L)\mu_{1,s,s}(L)\right\}\left(\frac{1-\pi}{1-\hat{\pi}}-1\right)\{p_{0,s}(L)-\hat{p}_{0,s}(L)\} \right]\\
    & + \mathbb{P}\left[\left\{\hat{p}_{0,s}(L)-p_{0,s}(L)\right\}\left(\frac{\pi}{\hat{\pi}}-1\right)\{p_{1,s}(L)\mu_{1,s,s}(L)-\hat{p}_{1,s}(L)\hat{\mu}_{1,s,s}(L)\}\right]\\
    &+\mathbb{P}\left[p_{1,s}(L)\mu_{1,s,s}(L)\left(\frac{1-\pi}{1-\hat{\pi}}-1\right)\{p_{0,s}(L)-\hat{p}_{0,s}(L)\} \right]\\
    & + \mathbb{P}\left[p_{0,s}(L)\left(\frac{\pi}{\hat{\pi}}-1\right)\{p_{1,s}(L)\mu_{1,s,s}(L)-\hat{p}_{1,s}(L)\hat{\mu}_{1,s,s}(L)\}\right]\\
    & +\mathbb{P}\left[ \{\hat{p}_{1,s}(L)\hat{\mu}_{1,s,s}(L)-p_{1,s}(L)\mu_{1,s,s}(L)\}\{p_{0,s}(L)-\hat{p}_{0,s}(L)\} \right]\\
    &= \left(\frac{1-\pi}{1-\hat{\pi}}-1\right)\mathbb{P}\left[ \left\{\hat{p}_{1,s}(L)\hat{\mu}_{1,s,s}(L)-p_{1,s}(L)\mu_{1,s,s}(L)\right\}\{p_{0,s}(L)-\hat{p}_{0,s}(L)\} \right]\\
    & + \left(\frac{\pi}{\hat{\pi}}-1\right)\mathbb{P}\left[\left\{\hat{p}_{0,s}(L)-p_{0,s}(L)\right\}\{p_{1,s}(L)\mu_{1,s,s}(L)-\hat{p}_{1,s}(L)\hat{\mu}_{1,s,s}(L)\}\right]\\
    &+\left(\frac{1-\pi}{1-\hat{\pi}}-1\right)\mathbb{P}\left[p_{1,s}(L)\mu_{1,s,s}(L)\{p_{0,s}(L)-\hat{p}_{0,s}(L)\} \right]\\
    & + \left(\frac{\pi}{\hat{\pi}}-1\right)\mathbb{P}\left[p_{0,s}(L)\{p_{1,s}(L)\mu_{1,s,s}(L)-\hat{p}_{1,s}(L)\hat{\mu}_{1,s,s}(L)\}\right]\\
    & +\mathbb{P}\left[ \{\hat{p}_{1,s}(L)\hat{\mu}_{1,s,s}(L)-p_{1,s}(L)\mu_{1,s,s}(L)\}\{p_{0,s}(L)-\hat{p}_{0,s}(L)\} \right].
\end{align*}
In the last step, we use that $A\indep L$ ($P(A=1|L)=\pi$).

Assuming $p_{1,s}(L)\mu_{1,s,s}(L)=O_p(1)$, $p_{0,s}(L)=O_p(1)$ and $0<\hat{\pi}<1$, we find the following rate conditions by applying the Cauchy-Schwarz inequality
\begin{align*}
    \mathbb{P}\left[\{p_{0,s}(L)-\hat{p}_{0,s}(L)\}^2 \right]^{1/2} &= o_p(1),\\
    \mathbb{P}\left[ \{p_1(L)\mu_{1,s,s}(L)-\hat{p}_{1,s}(L)\hat{\mu}_{1,s,s}(L)\}^2\right]^{1/2} &= o_p(1)\\
    \mathbb{P}\left[\{p_{0,s}(L)-\hat{p}_{0,s}(L)\}^2 \right]^{1/2}\mathbb{P}\left[ \{p_{1,s}(L)\mu_{1,s,s}(L)-\hat{p}_{1,s}(L)\hat{\mu}_{1,s,s}(L)\}^2\right]^{1/2} &= o_p(n^{-1/2}).
\end{align*}

Consequently, we have the following rate conditions.
\begin{align*}
    &\sum_{s=0}^t\mathbb{P}\left[\left\{p_{a,s}(L)-\hat{p}_{a,s}(L)\right\}^2\right]^{1/2}=o_p(1),\\
    &\sum_{s=0}^t\mathbb{P}\left[\{p_{a,s}(L)\mu_{a,s,s}(L)-\hat{p}_{a,s}(L)\hat{\mu}_{a,s,s}(L)\}^2\right]^{1/2}=o_p(1),\\
    &\sum_{s=0}^t\mathbb{P}\left[\{p_{a,s-1}(L)\mu_{a,s,s-1}(L)-\hat{p}_{a,s-1}(L)\hat{\mu}_{a,s,s-1}(L)\}^2\right]^{1/2}=o_p(1),\\
    &\sum_{s=0}^t\mathbb{P}\left[\{p_{1-a,s}(L)-\hat{p}_{1-a,s}(L)\}^2 \right]^{1/2}\mathbb{P}\left[ \{p_{a,s}(L)\mu_{a,s,s}(L)-\hat{p}_{a,s}(L)\hat{\mu}_{a,s,s}(L)\}^2\right]^{1/2} = o_p(n^{-1/2}),\\
    &\sum_{s=0}^t\mathbb{P}\left[\{p_{1-a,s-1}(L)-\hat{p}_{1-a,s-1}(L)\}^2 \right]^{1/2}\mathbb{P}\left[ \{p_{a,s-1}(L)\mu_{a,s,s-1}(L)-\hat{p}_{a,s-1}(L)\hat{\mu}_{a,s,s-1}(L)\}^2\right]^{1/2} = o_p(n^{-1/2}).
\end{align*}

Under these rate assumptions, we have that 
\[
\sqrt{n}\left(\hat{\Phi}^*_t-\Phi_t\right) = \frac{1}{n}\sum_{i=1}^n\sum_{s=0}^t \zeta^*_{s,s-1}(O_i) - I(t>s)\zeta^*_{s,s}(O_i) - \Phi_t + o_p(1),
\]
which converges to a mean zero normal distribution with variance equal to the variance of (\ref{CPLOTestEIFsp}) by the central limit theorem.

The rate condtions can be achieved, for instance, when both estimators of the nuisance functions in the product converge at a rate faster than $n^{-1/4}$, or when one is estimated at parametric rate while the other is merely consistent. This compensation in convergence rates is, however, slightly more restrictive here than in the case of AIPW estimators of the ATE.
The reason is that both factors in the product involve $p_{a,s}(L)$ and $\hat{p}_{a,s}(L)$, making it less plausible that one can converge quickly when the other converges slowly. 
Nevertheless, since $\mathbb{P}\left[ \{p_{a,s}(L)\mu_{a,s,s}(L)-\hat{p}_{a,s}(L)\hat{\mu}_{a,s,s}(L)\}^2\right]^{1/2}$ can be upper bounded by
\begin{align}\label{uppboundrate}
\begin{split}
    &\mathbb{P}\left[ p_{a,s}(L)^2\{\mu_{a,s,s}(L)-\hat{\mu}_{a,s,s}(L)\}^2\right]^{1/2} + \mathbb{P}\left[ \mu_{a,s,s}(L)^2\{p_{a,s}(L)-\hat{p}_{a,s}(L)\}^2\right]^{1/2}\\
    &+ \mathbb{P}\left[ (\{p_{a,s}(L)-\hat{p}_{a,s}(L)\}\{\hat{\mu}_{a,s,s}(L)-\mu_{a,s,s}(L)\})^2\right]^{1/2},    
\end{split}
\end{align}
meaningful forms of rate double robustness can still arise if $\mu_{a,s,s}(L)=0$ almost surely. In that case, the first term in (\ref{uppboundrate}) dominates and condition 4 in Theorem \ref{th: CPLOT} simplifies to $$\sum_{s=0}^t\mathbb{P}\left[\{p_{1-a,s}(L)-\hat{p}_{1-a,s}(L)\}^2 \right]^{1/2}\mathbb{P}\left[ p_{a,s}(L)^2\{\mu_{a,s,s}(L)-\hat{\mu}_{a,s,s}(L)\}^2\right]^{1/2} = o_p(n^{-1/2}).$$

This result extends naturally to observational studies, provided that baseline covariates $L$ suffice to adjust for confounding of the effects of treatment on both the outcome and time to ICE. The estimators remain essentially unchanged, but the rate requirements would be that all of the following terms are $o_p(n^{-1/2})$.
\begin{align*}
    &\sum_{s=0}^t\mathbb{P}\left[\left\{\pi(L)-\hat{\pi}(L)\right\}^2 \right]^{1/2}\mathbb{P}\left[\left\{p_{a,s}(L)-\hat{p}_{a,s}(L)\right\}^2\right]^{1/2},\\
    &\sum_{s=0}^t\mathbb{P}\left[\left\{\pi(L)-\hat{\pi}(L)\right\}^2 \right]^{1/2}\mathbb{P}\left[\{p_{a,s}(L)\mu_{a,s,s}(L)-\hat{p}_{a,s}(L)\hat{\mu}_{a,s,s}(L)\}^2\right]^{1/2},\\
    &\sum_{s=0}^t\mathbb{P}\left[\left\{\pi(L)-\hat{\pi}(L)\right\}^2 \right]^{1/2}\mathbb{P}\left[\{p_{a,s-1}(L)\mu_{a,s,s-1}(L)-\hat{p}_{a,s-1}(L)\hat{\mu}_{a,s,s-1}(L)\}^2\right]^{1/2},\\
    &\sum_{s=0}^t\mathbb{P}\left[\left\{\pi(L)-\hat{\pi}(L)\right\}\{p_{1-a,s}(L)-\hat{p}_{1-a,s}(L)\}\{p_{a,s}(L)\mu_{a,s,s}(L)-\hat{p}_{a,s}(L)\hat{\mu}_{a,s,s}(L)\}\right],\\
    &\sum_{s=0}^t\mathbb{P}\left[\left\{\pi(L)-\hat{\pi}(L)\right\}\{p_{1-a,s-1}(L)-\hat{p}_{1-a,s-1}(L)\}\{p_{a,s-1}(L)\mu_{a,s,s-1}(L)-\hat{p}_{a,s-1}(L)\hat{\mu}_{a,s,s-1}(L)\}\right],
\end{align*}
where $\pi(L) = P(A=1|L)$.

\newpage
\section*{Appendix C: Simulation study details}
\subsection*{More details about the alternative methods}
We applied the SACE estimator as proposed by \cite{hayden2005estimator} and implemented in \cite{vonfelten2025}. Survival probabilities were estimated using logistic regression models with main effects for all baseline covariates, including the baseline outcome. The IPCW estimator gives an estimate for the hypothetical estimand for the treatment effect had there been no intercurrent events. It was obtained as the treatment coefficient from a generalized linear model including the same covariates, fitted among survivors and weighted by the inverse of the estimated survival probabilities. These probabilities were derived from a logistic regression with main effects of all baseline covariates. The LOCF method compared the last observed outcomes of all individuals using an unpaired t-test. The naive survivors method restricted this comparison to outcomes among survivors only.

\subsection*{Table to visualize the distribution of $T$ in each treatment group}
Table \ref{tab: distT} visualizes the distribution of $T$ in each treatment group. This is done by generating a very large sample ($n=10^6$) and then calculating for each treatment arm which percentage has a certain value of $T\in\{0,1,\ldots,\tau\}$. 

\begin{table}[h!]
\centering
\begin{tabular}{ccccccc}
  & \multicolumn{2}{c}{Setting 1} &  \multicolumn{2}{c}{Setting 2} & \multicolumn{2}{c}{Setting 3} \\ 
T & $A=0$ & $A=1$ & $A=0$ & $A=1$ & $A=0$ & $A=1$  \\ 
  0 & 6  & 2 & 0    & 0    & 6  & 1  \\
  1 & 8  & 3 & 10   & 18   & 8  & 2  \\
  2 & 7  & 3 & 9    & 22   & 8  & 2  \\
  3 & 7  & 3 & 8    & 23   & 8  & 2  \\
  4 & 6  & 3 & 8    & 19   & 7  & 3  \\
  5 & 67 & 85& 64   & 18   & 64 & 90 \\
\end{tabular}
\caption{Visualization of the distribution of $T$ in each treatment arm and for each setting, expressed in percentages.}
\label{tab: distT}
\end{table}

\subsection*{Additional simulation results}
\subsubsection*{Observational study}
To evaluate the finite-sample performance of the proposed methods under the null of no treatment effect on the outcome in an observational study, we conducted a simulation experiment with 1000 runs, where we set $\tau=5$ and sample size $n=250$. Data is generated as in setting 1, except that $A$ is not randomized. Instead, we generated a binary $A$ according to 
\[
    P(A=1|L) = \expit(0.5+L_4/3+L_5/4+L_6/5-L_8/3-L_9/4-L_{10}/5).
\]
Propensity scores were estimated using Super Learner (with the same library as for the other nuisance parameters), and the remaining nuisance parameters were estimated as in the other simulation experiments.

Table \ref{tab: observational} presents the simulation results for this setting. The unadjusted PLOT estimator, SACE, LOCF and the survivor-based estimators are now biased, as they do not account for confounding. IPCW is slightly biased and standard errors are underestimated, leading to (significant) undercoverage. In contrast, the (C)PLOT estimators are approximately unbiased with 95\% confidence intervals having coverage near the nominal level, except when no cross-fitting is used. The high variablity in the estimators with cross-fitting is due to one outlier. 

\begin{table}[ht]
\centering
\begin{tabular}{lcccc}
  \hline
 & Estimate  & SD  & SE & Coverage \\ 
  \hline
PLOTunadj & -0.507 (-0.516) & 0.486 (0.434) & 0.485 (0.484) & 83.2 \\ 
  PLOTadj & -0.095 (-0.081) & 0.413 (0.216) & 0.145 (0.127) & 67.9 \\ 
  PLOTadj-CF & 0.486 (-0.090) & 18.290 (0.223) & 0.710 (0.225) & 92.5 \\ 
  CPLOT & -0.051 (-0.049) & 0.270 (0.223) & 0.139 (0.128) & 68.7 \\ 
  CPLOT-CF & 0.478 (-0.042) & 16.270 (0.229) & 0.661 (0.224) & 93.2 \\ 
  SACE & -0.440 (-0.449) & 0.491 (0.432) & 1.41$\times 10^7$ (0.636) & 93.2 \\ 
  IPCW & 0.071 (0.067) & 0.243 (0.222) & 0.222 (0.221) & 91.3 \\ 
  Survivors & -0.363 (-0.371) & 0.536 (0.489) & 0.533 (0.528) & 89.1 \\ 
  LOCF & -0.342 (-0.353) & 0.495 (0.446) & 0.494 (0.493) & 90.2 \\  
   \hline
\end{tabular}
\caption{Simulation results for an observational study. Estimate: mean (median) of estimates; SD: standard deviation of estimates (robust sd using \texttt{cov.mcd} function in R); SE: mean (median) of estimated standard errors; Cov: coverage of 95\% confidence intervals. PLOTunadj denotes unadjusted estimator (\ref{UnadjEst}). PLOTadj(-CF) denotes adjusted estimator (\ref{AdjDMLsp}) (with 5-fold cross-fitting). CPLOT(-CF) denotes estimator (\ref{CondDMLsp}) (with 5-fold cross-fitting). SACE is the estimator proposed by \cite{hayden2005estimator}. IPCW is an inverse probability of censoring weighting estimator. `Survivors' is the treatment effect among individuals reaching the end of the study. LOCF is the last observation carried forward estimator.}
\label{tab: observational}
\end{table}

\subsubsection*{Additional simulation results for Section \ref{sec: assumptions}}
For illustrative purposes, Section \ref{sec: assumptions} considered the following outcome model:
\begin{equation*}
    Y(t)=\alpha(t)+\beta_0 L+\beta_1 Lt+\gamma_0 U + \gamma_1 Ut + \epsilon_t.
\end{equation*}
Here, we report additional simulation results (based on 1000 Monte Carlo replicates) generated under this model with $\epsilon_t\sim N(0,1)$, $T=\min(\lfloor T_{ICE}\rfloor,\tau)$ and the time to ICE, $T_{ICE}$, is generated from a Weibull distribution with shape parameter 1.5 and scale parameter equal to $\exp(2.5+0.8A+L+0.2(2A-1)U)$. The baseline covariates $L$ and $U$ are independent standard normal variables. We fix $\beta_0=\beta_1=\gamma_0=1$, $\tau=5$, $n=250$ and vary $\gamma_1\in\{-2,-1.5,-1,-0.5,0,0.5,1,1.5,2\}$ to examine performance when the proposed estimands are expected to deviate from zero. The values for $\gamma_1$ were selected to span a range of dependence strengths between the outcome process and the unmeasured variable $U$, as quantified by the partial $R^2$,
\[
    \frac{1}{\tau+1}\sum_{t=0}^\tau\frac{E(\var(Y(t)|L))-E(\var(Y(t)|L,U))}{E(\var(Y(t)|L))} = \frac{1}{\tau+1}\sum_{t=0}^\tau\frac{(1+\gamma_1 t)^2}{(1+\gamma_1 t)^2+1},
\]
with corresponding values reported in Table \ref{tab:R2}. 
\begin{table}[h!]
    \centering
    \begin{tabular}{c|ccccccccc}
        $\gamma_1$ & -2 & -1.5 & -1 & -0.5 & 0 & 0.5 & 1 & 1.5 & 2 \\
        $R^2_{\text{partial}}$ & 0.80 & 0.73 & 0.61 & 0.35 & 0.50 & 0.78 & 0.85 & 0.87 & 0.89
    \end{tabular}
    \caption{Partial $R^2$ values corresponding to each choice for $\gamma_1$ in the outcome model.}
    \label{tab:R2}
\end{table}

\begin{figure}[h!]
    \centering
    \includegraphics[width=0.9\linewidth]{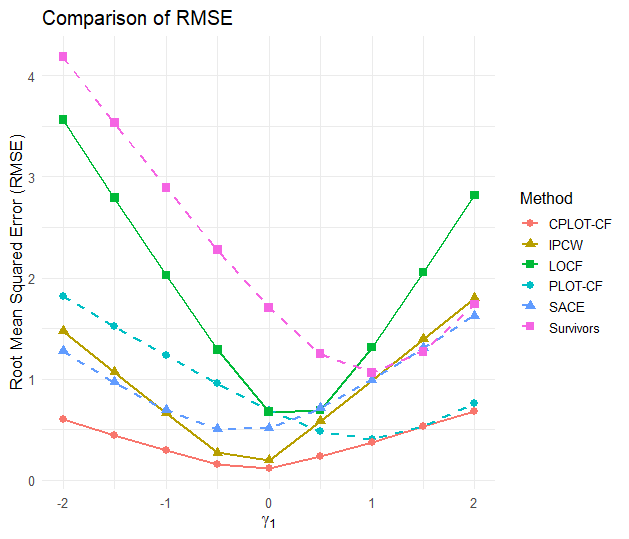}
    \caption{Comparison of RMSE for different methods for $\gamma_1\in\{-2,-1.5,-1,-0.5,0,0.5,1,1.5,2\}$.}
    \label{fig: rmse}
\end{figure}
\begin{figure}[h!]
    \centering
    \begin{subfigure}[b]{0.48\linewidth}
        \centering
        \includegraphics[width=\linewidth]{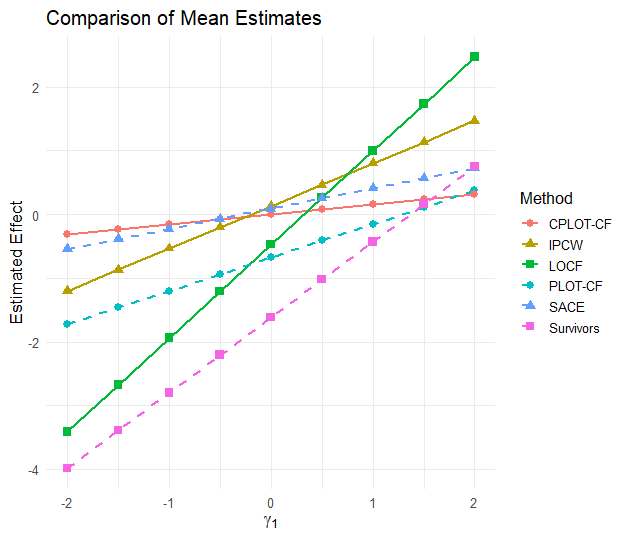}
    \end{subfigure}
    \hfill
    \begin{subfigure}[b]{0.48\linewidth}
        \centering
        \includegraphics[width=\linewidth]{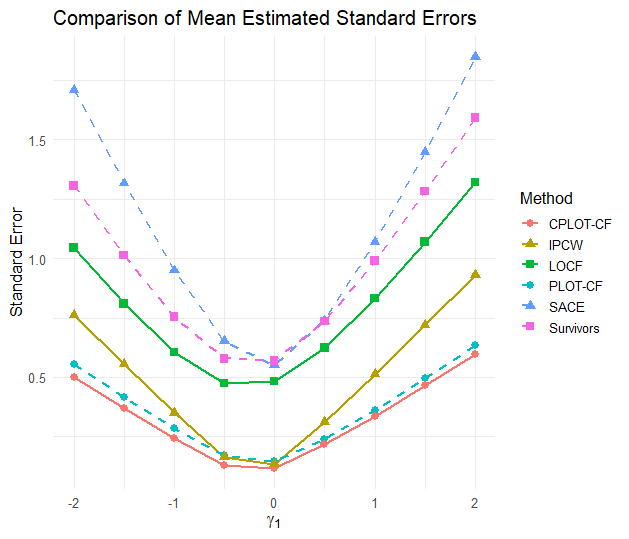}
    \end{subfigure}
    \caption{Comparison of mean estimate (left) and mean standard error (right) for different methods for $\gamma_1\in\{-2,-1.5,-1,-0.5,0,0.5,1,1.5,2\}$.}
    \label{fig:combined}
\end{figure}

Each scenario compares our estimators (with 5-fold cross-fitting) against a SACE estimator, an IPCW estimator, a LOCF estimator and a naive survivors-only estimator that ignores baseline covariates. Figure \ref{fig: rmse} presents the root mean squared error (RMSE) for each method across values of $\gamma_1$, and Figure \ref{fig:combined} displays the mean estimates (left) and mean standard errors (right). 

The results show clear deviations from zero for LOCF, Survivors, IPCW and PLOT-CF, whereas SACE and CPLOT-CF remain close to zero across all values of $\gamma_1$. Deviations grow as $\gamma_1$ moves away from zero, but CPLOT-CF remains closest to zero across all settings. In terms of mean standard errors, IPCW and our two proposed estimators perform best, while SACE, LOCF and Survivors show high variability, particularly for larger $|\gamma_1|$. These patterns are reflected in the RMSE results, where CPLOT-CF achieves the most favorable overall performance.

\newpage
\section*{Appendix D: Estimators for (log) ratio estimands (\ref{PLOTratio}) and (\ref{CPLOTratio})}
Below, we show how to use the same strategies as in Section \ref{sec: identest} to estimate
\begin{equation}\label{PLOTlogratio}
\log\frac{E\left\{Y^1(\min(T^1, T^{*0}, t))\right\}}{E\left\{Y^{*0}(\min(T^1, T^{*0}, t))\right\}}
\end{equation}
and
\begin{equation}\label{CPLOTlogratio}
\log\frac{E\left[E \left\{Y^1(\min(T^1, T^{*0}, t))|L=L^*\right\}\right]}{E\left[E \left\{Y^{*0}(\min(T^1, T^{*0}, t))|L=L^*\right\}\right]}.
\end{equation}

\subsection*{Simple PLOT estimator}
An unadjusted estimator for (\ref{PLOTlogratio}) is 
\begin{align*}
    \log\frac{\sum_{s=0}^t\frac{1}{\sum_{i,j} A_i(1-A_j)}\sum_{i,j} Y_i(s)A_i(1-A_j)I\left\{\min(T_i, T_j, t) = s\right\}}{\sum_{s=0}^t\frac{1}{\sum_{i,j} A_i(1-A_j)}\sum_{i,j} Y_j(s)A_i(1-A_j)I\left\{\min(T_i, T_j, t) = s\right\}}.
\end{align*}
The standard error can be computed via the bootstrap or as explained below with $L=\emptyset$.

\subsection*{Efficient PLOT estimator}
Following the same reasoning and notation as in the main paper and using the Delta method, we find that the EIF of $\Psi_t^{LR} = \log\frac{\Psi_t^{LR,1}}{\Psi_t^{LR,0}}$, where 
\begin{align*}
    \Psi_t^{LR,1} &= E\left\{Y^1(\min(T^1, T^{*0}, t))\right\}, \\
    \Psi_t^{LR,0} &= E\left\{Y^{*0}(\min(T^1, T^{*0}, t))\right\},
\end{align*}
under $\mathcal{P}_\pi$ equals
\begin{align}\label{PLOTlogratioEIF}
    \frac{\sum_{s=0}^t \xi^{*LR,1}_{s,s-1}(O) - I(t>s)\xi^{*LR,1}_{s,s}(O)}{\Psi_t^{LR,1}}- \frac{\sum_{s=0}^t \xi^{*LR,0}_{s,s-1}(O) - I(t>s)\xi^{*LR,0}_{s,s}(O)}{\Psi_t^{LR,0}},
\end{align}
where
\begin{align*}
    \xi^{*LR,1}_{s,u}(O) &= \varphi^*_{\eta_{0,u}}(O)\gamma_{1,s,u} + \eta_{0,u}\varphi^*_{\gamma_{1,s,u}}(O),\\
    \xi^{*LR,0}_{s,u}(O) &= \varphi^*_{\eta_{1,u}}(O)\gamma_{0,s,u} -\eta_{1,u}\varphi^*_{\gamma_{0,s,u}}(O).
\end{align*}
The resulting estimator for $\Psi_t^{LR}$ is $\log\hat{\Psi}_t^{*LR,1}-\log\hat{\Psi}_t^{*LR,0}$, where (for $a=0,1$)
\begin{equation*}
    \hat{\Psi}^{*LR,a}_t = \frac{1}{2n}\sum_{i=1}^n\sum_{s=0}^t\hat{\xi}^{*LR,a}_{s,s-1}(O_i) - I(t>s)\hat{\xi}^{*LR,a}_{s,s}(O_i).
\end{equation*}
Its variance can be consistently estimated as 1 over $n$ times the sample variance of (\ref{PLOTlogratioEIF}).

\subsection*{CPLOT estimator}
Following the same reasoning and notation as in the main paper and using the Delta method, we find that the EIF of $\Phi_t^{LR} = \log\frac{\Phi_t^{LR,1}}{\Phi_t^{LR,0}}$, where 
\begin{align*}
    \Phi_t^{LR,1} &= E\left[E \left\{Y^1(\min(T^1, T^{*0}, t))|L=L^*\right\}\right],\\
    \Phi_t^{LR,0} &= E\left[E \left\{Y^{*0}(\min(T^1, T^{*0}, t))|L=L^*\right\}\right],
\end{align*}
under $\mathcal{P}_\pi$ equals
\begin{align}\label{CPLOTlogratioEIF}
    \frac{\sum_{s=0}^t \zeta^{*LR,1}_{s,s-1}(O) - I(t>s)\zeta^{*LR,1}_{s,s}(O)}{\Phi_t^{LR,1}}- \frac{\sum_{s=0}^t \zeta^{*LR,0}_{s,s-1}(O) - I(t>s)\zeta^{*LR,0}_{s,s}(O)}{\Phi_t^{LR,0}},
\end{align}
where
\begin{align*}
    \zeta^{*LR,a}_{s,u}(O) &= p_{1,u}(L)p_{0,u}(L)\mu_{a,s,u}(L) + \frac{I(A=1-a)}{P(A=1-a)}\left\{I(T>u) - p_{1-a,u}(L)\right\}p_{a,u}(L)\mu_{a,s,u}(L)\\
    &\qquad + p_{1-a,u}(L)\frac{I(A=a)}{P(A=a)}\left\{Y(s)I(T>u) - p_{a,u}(L)\mu_{a,s,u}(L)\right\}.
\end{align*}
The resulting estimator for $\Phi_t^{LR}$ is $\log\hat{\Phi}_t^{*LR,1}-\log\hat{\Phi}_t^{*LR,0}$, where (for $a=0,1$)
\begin{equation*}
    \hat{\Phi}^{*LR,a}_t = \frac{1}{n}\sum_{i=1}^n\sum_{s=0}^t\hat{\zeta}^{*LR,a}_{s,s-1}(O_i) - I(t>s)\hat{\zeta}^{*LR,a}_{s,s}(O_i).
\end{equation*}
Its variance can be consistently estimated as 1 over $n$ times the sample variance of (\ref{CPLOTlogratioEIF}).

\newpage
\section*{Appendix E: Proofs of statements in Section~\ref{sec: assumptions}}
\subsection*{Evaluating the estimands under the example model}
Consider a continuous outcome $Y(t)=\alpha(t)+\beta_0 L+\beta_1 Lt+\gamma_0 U + \gamma_1 Ut + \epsilon_t$, where $L$ and $U$ are baseline covariates but only $L$ is measured, and $\epsilon_t$ is a mean zero random error term. Clearly, there is no treatment effect on the outcome. However, we assume that $A$, $L$ and $U$ all influence the timing of ICEs (i.e., $T$). Under this model, $\Psi_t$ reduces to 
\begin{align*}
    & \sum_{s=0}^tE\left\{Y(s)-Y^*(s)|\min(T^1,T^{*0},t)=s\right\}P\left\{\min(T^1,T^{*0},t)=s\right\}\\
    &= \beta_0E(L-L^*) + \beta_1 E\left\{(L-L^*)\min(T^1,T^{*0},t)\right\} + \gamma_0E(U-U^*) + \gamma_1 E\left\{(U-U^*)\min(T^1,T^{*0},t)\right\}\\
    &= \beta_1 E\left\{(L-L^*)\min(T^1,T^{*0},t)\right\} + \gamma_1 E\left\{(U-U^*)\min(T^1,T^{*0},t)\right\},
\end{align*}
while $\Phi_t$ (conditioning on only $L$) reduces to
\begin{align*}
    & \sum_{s=0}^tE\left[E\left\{Y(s)-Y^*(s)|\min(T^1,T^{*0},t)=s,L=L^*\right\}P\left\{\min(T^1,T^{*0},t)=s|L=L^*\right\}\right]\\
    &= \beta_0 E\{E(L-L^*|L=L^*)\} + \beta_1 E\left[E\left\{(L-L^*)\min(T^1,T^{*0},t)|L=L^*\right\}\right]\\
    &\qquad + \gamma_0E\{E(U-U^*|L=L^*)\} + \gamma_1 E\left[E\left\{(U-U^*)\min(T^1,T^{*0},t)|L=L^*\right\}\right]\\
    &= \gamma_1 E\left[E\left\{(U-U^*)\min(T^1,T^{*0},t)|L=L^*\right\}\right].
\end{align*}

\subsection*{Null validity of $\Phi_t$ under various assumptions}
We first show that $\Phi_t$ is zero under the null when condition \ref{ass1:ipw} of Assumption \ref{ass1} holds.
\begin{align*}
    & E\left[E\left\{Y^1 (\min(T^1, T^{*0}, t))|L=L^*\right\}\right]\\
    &= \sum_{s=0}^t E\left[E\left\{Y^1(s)|T^1>s-1,L\right\}P(T^1>s-1|L)P(T^0>s-1|L)\right]\\
    &\qquad - I(t>s)E\left[E\left\{Y^1(s)|T^1>s,L\right\}P(T^1>s|L)P(T^0>s|L)\right]\\
    &= \sum_{s=0}^t E\left[E\left\{Y^1(s)|L\right\}P(T^1>s-1|L)P(T^0>s-1|L)\right]\\
    &\qquad - I(t>s)E\left[E\left\{Y^1(s)|L\right\}P(T^1>s|L)P(T^0>s|L)\right]\\
    &= \sum_{s=0}^t E\left[E\left\{Y^0(s)|L\right\}P(T^1>s-1|L)P(T^0>s-1|L)\right]\\
    &\qquad - I(t>s)E\left[E\left\{Y^0(s)|L\right\}P(T^1>s|L)P(T^0>s|L)\right],
\end{align*}
where the second equation holds under condition \ref{ass1:ipw} and the last equation under the null. Therefore, we have that $\Phi_t=0$ under the null.

Next, we show that $\Phi_t$ is zero under the null when condition \ref{ass1:yt-y0} holds.
\begin{align*}
    &E\left\{Y^1(\min(T^1, T^{*0}, t)) - Y^{*0}(\min(T^1, T^{*0}, t))|L=L^*=l\right\}\\
    &=E\left\{Y^1(\min(T^1, T^{*0}, t)) - Y(0) - Y^{*0}(\min(T^1, T^{*0}, t)) + Y^*(0)|L=L^*=l\right\}\\
    &= \iiint \delta f_{Y^1(\min(v, w, t)) - Y(0)|T^1=v,T^{*0}=w,L=l,L^*=l}(\delta)f_{T^1|L=l}(v)f_{T^{*0}|L^*=l}(w)\,\mathrm{d}\delta\mathrm{d}v\mathrm{d}w\\
    &\quad - \iiint \delta f_{Y^{*0}(\min(v, w, t)) - Y^*(0)|T^1=v,T^{*0}=w,L=l,L^*=l}(\delta)f_{T^1|L=l}(v)f_{T^{*0}|L^*=l}(w)\,\mathrm{d}\delta\mathrm{d}v\mathrm{d}w\\
    &= \iiint \delta f_{Y^1(\min(v, w, t)) - Y(0)|T^1=v,L=l}(\delta)f_{T^1|L=l}(v)f_{T^{0}|L=l}(w)\,\mathrm{d}\delta\mathrm{d}v\mathrm{d}w\\
    &\quad - \iiint \delta f_{Y^{*0}(\min(v, w, t)) - Y^*(0)|T^{*0}=w,L^*=l}(\delta)f_{T^1|L=l}(v)f_{T^{0}|L=l}(w)\,\mathrm{d}\delta\mathrm{d}v\mathrm{d}w\\
    &= \iiint \delta f_{Y^1(\min(v, w, t)) - Y(0)|L=l}(\delta)f_{T^1|L=l}(v)f_{T^{0}|L=l}(w)\,\mathrm{d}\delta\mathrm{d}v\mathrm{d}w\\
    &\quad - \iiint \delta f_{Y^{0}(\min(v, w, t)) - Y(0)|L=l}(\delta)f_{T^1|L=l}(v)f_{T^{0}|L=l}(w)\,\mathrm{d}\delta\mathrm{d}v\mathrm{d}w\\
    &=0,
\end{align*}
where the fourth equation holds under condition \ref{ass1:yt-y0} and the last equation under the null.

Finally, we show that $\Phi_t$ is zero under the null $Y^1(t)=Y^0(t),\forall t\leq\min(T^1, T^0)$ when Assumption \ref{ass: sace} holds. In that case, we have that
\begin{align*}
    & E\left[E\left\{Y^1 (\min(T^1, T^{*0}, t))|L=L^*\right\}\right]\\
    &= \sum_{s=0}^t E\left[E\left\{Y^1(s)|T^1>s-1,L\right\}P(T^1>s-1|L)P(T^0>s-1|L)\right]\\
    &\qquad - I(t>s)E\left[E\left\{Y^1(s)|T^1>s,L\right\}P(T^1>s|L)P(T^0>s|L)\right]\\
    &= \sum_{s=0}^t E\left[E\left\{Y^1(s)|T^1>s-1,T^0>s-1,L\right\}P(T^1>s-1|L)P(T^0>s-1|L)\right]\\
    &\qquad - I(t>s)E\left[E\left\{Y^1(s)|T^1>s,T^0>s,L\right\}P(T^1>s|L)P(T^0>s|L)\right]\\
    &= \sum_{s=0}^t E\left[E\left\{Y^0(s)|T^1>s-1,T^0>s-1,L\right\}P(T^1>s-1|L)P(T^0>s-1|L)\right]\\
    &\qquad - I(t>s)E\left[E\left\{Y^0(s)|T^1>s,T^0>s,L\right\}P(T^1>s|L)P(T^0>s|L)\right],
\end{align*}
where the second equation holds under Assumption \ref{ass: sace} and the last equation under the null. Therefore, we have that $\Phi_t=0$ under the null.

An alternative choice for the null hypothesis is 
\begin{align*}
    \begin{cases}
        &E\left\{Y^1(t)|T^1\geq t,T^0\geq t,L\right\} = E\left\{Y^0(t)|T^1\geq t,T^0\geq t,L\right\},\;\forall t\\
        &E\left\{Y^1(t)|T^1> t,T^0> t,L\right\} = E\left\{Y^0(t)|T^1> t,T^0> t,L\right\},\;\forall t
    \end{cases}.
\end{align*}

\newpage
\section*{Appendix F: Extension of results in Section~\ref{sec: connection} to ratio estimand \eqref{CPLOTratio} and proofs}

Extension of the results in Section~\ref{sec: connection} to the ratio estimand \eqref{CPLOTratio} relies on the following assumption for an arbitrary, time-specific hypothetical effect $\psi_s$:
\begin{equation}\label{interpretGeneralratio}
E(Y^1(s) | L) = E(Y^0(s) | L)e^{\psi_s}, \quad \forall s \in \{0, 1, \ldots, \tau\}
\end{equation}
where $\psi_0 = 0$. Again, $\psi_s$ must be interpreted as the average treatment effect at time $s$ had all individuals remained free of ICEs. Analogously to Section \ref{sec: assumptions}, we define the vector $\psi = (\psi_1, \ldots, \psi_\tau) \in \mathbb{R}^{\tau}$ as the solution to:
\begin{align*}
    &E\left[ E\left\{ Y^1(M_t)e^{-\psi_{M_t}} - Y^{*0}(M_t) | L=L^* \right\} \right] = 0, \quad \forall t \in \{1, \ldots, \tau\}\\
    \iff & E\left[ E\left\{ Y^1(M_t)e^{-\psi_{M_t}} | L=L^* \right\} \right] = E\left[E\left\{Y^{*0}(M_t)|L=L^*\right\}\right], \quad \forall t \in \{1, \ldots, \tau\},
\end{align*}
where $M_t = \min(T^1, T^{*0}, t)$. The solution (with proof given below) of the system equations can be expressed as:
\begin{align*}
    e^{-\psi_t} &= \frac{\Phi_{0,t} - \Phi_{0,t-1}}{E\left[E\left\{Y^1(t)\mathrm{I}(\min(T^1,T^{*0},t)=t)|L=L^*\right\}\right]}\\
    &\qquad + e^{-\psi_{t-1}}\frac{E\left[E\left\{Y^1(t-1)\mathrm{I}(\min(T^1,T^{*0},t)=t)|L=L^*\right\}\right]}{E\left[E\left\{Y^1(t)\mathrm{I}(\min(T^1,T^{*0},t)=t)|L=L^*\right\}\right]}\\
    & = \prod_{j=1}^{t} \left( \frac{E\left[E\left\{Y^1(j-1)\mathrm{I}(\min(T^1,T^{*0},j)=j)|L=L^*\right\}\right]}{E\left[E\left\{Y^1(j)\mathrm{I}(\min(T^1,T^{*0},j)=j)|L=L^*\right\}\right]} \right) \\
    &\qquad + \sum_{k=1}^{t} \left[ \frac{\Phi_{0,k} - \Phi_{0,k-1}}{E\left[E\left\{Y^1(k)\mathrm{I}(\min(T^1,T^{*0},k)=k)|L=L^*\right\}\right]}\right.\\
    &\qquad\qquad\times\left. \prod_{j=k+1}^{t} \left( \frac{E\left[E\left\{Y^1(j-1)\mathrm{I}(\min(T^1,T^{*0},j)=j)|L=L^*\right\}\right]}{E\left[E\left\{Y^1(j)\mathrm{I}(\min(T^1,T^{*0},j)=j)|L=L^*\right\}\right]} \right) \right],
\end{align*}
where
\[
    \Phi_{0,t} = E\left[E\left\{Y^{*0}(\min(T^1, T^{*0}, t))|L=L^*\right\}\right].
\]
This general approach provides an interpretation of $\hat{\psi}_t$ as the hypothetical treatment effect at time $t$ had there been no ICEs.

\subsection{Proofs}
For the additive estimand, we solve the following system of equations:
\begin{align*}
    &E\left[E\left\{Y^1(\min(T^1, T^{*0}, t)) - Y^{*0}(\min(T^1, T^{*0}, t))|L=L^*\right\}\right]\\
    &= E\left[ E\left\{\psi_{\min(T^1,T^{*0},t)}|L=L^*\right\}\right],\quad \forall t\in\{1,\ldots,\tau\}.
\end{align*}
For an arbitrary $t\in\{1,\ldots,\tau\}$, we have
\begin{align*}
&\begin{cases}
    \Phi_t = E\left[E\left\{\psi_{\min(T^1,T^{*0},t)}|L=L^*\right\}\right] = \sum_{s=0}^t\psi_sE\left[P\left\{\min(T^1,T^{*0},t)=s|L=L^*\right\}\right]\\
    \Phi_{t-1} = E\left[E\left\{\psi_{\min(T^1,T^{*0},t-1)}|L=L^*\right\}\right] = \sum_{s=0}^{t-1}\psi_sE\left[P\left\{\min(T^1,T^{*0},t-1)=s|L=L^*\right\}\right].
\end{cases}
\end{align*}
Consequently,
\begin{align*}
    \Phi_t - \Phi_{t-1} &= \psi_tE\left[P\left\{\min(T^1,T^{*0},t)=t|L=L^*\right\}\right]\\
    &\quad + \psi_{t-1}E\left[P\left\{\min(T^1,T^{*0},t)=t-1|L=L^*\right\}-P\left\{\min(T^1,T^{*0},t-1)=t-1|L=L^*\right\}\right]\\
    &= (\psi_t-\psi_{t-1})E\left[P\left\{\min(T^1,T^{*0},t)=t|L=L^*\right\}\right].
\end{align*}
We find that
\begin{align*}
    \psi_t &= \psi_{t-1} + \frac{\Phi_t - \Phi_{t-1}}{E\left[P\left\{\min(T^1,T^{*0},t)=t|L=L^*\right\}\right]}\\
    &= \sum_{s=1}^t\frac{\Phi_s - \Phi_{s-1}}{E\left[P\left\{\min(T^1,T^{*0},s)=s|L=L^*\right\}\right]}.
\end{align*}

For the ratio estimand, we solve the following system of equations:
\begin{align*}
    &E\left[E\left\{Y^1(\min(T^1, T^{*0}, t))e^{-\psi_{\min(T^1,T^{*0},t)}}|L=L^*\right\}\right]\\
    &= E\left[ E\left\{Y^{*0}(\min(T^1, T^{*0}, t))|L=L^*\right\}\right],\quad \forall t\in\{1,\ldots,\tau\}.
\end{align*}
For an arbitrary $t\in\{1,\ldots,\tau\}$, we have
\begin{align*}
&\begin{cases}
    \Phi_{0,t} = E\left[E\left\{Y^1(\min(T^1, T^{*0}, t))e^{-\psi_{\min(T^1, T^{*0}, t)}}|L=L^*\right\}\right] \\
    \Phi_{0,t-1} = E\left[E\left\{Y^1(\min(T^1, T^{*0}, t-1))e^{-\psi_{\min(T^1, T^{*0}, t-1)}}|L=L^*\right\}\right] \\
\end{cases}\\
\iff &\begin{cases}
    \Phi_{0,t} = \sum_{s=0}^te^{-\psi_s}E\left[E\left\{Y^1(s)\mathrm{I}(\min(T^1,T^{*0},t)=s)|L=L^*\right\}\right] \\
    \Phi_{0,t-1} = \sum_{s=0}^{t-1}e^{-\psi_s}E\left[E\left\{Y^1(s)\mathrm{I}(\min(T^1,T^{*0},t-1)=s)|L=L^*\right\}\right]
\end{cases}.
\end{align*}
Consequently,
\begin{align*}
    \Phi_{0,t} - \Phi_{0,t-1} &= e^{-\psi_t}E\left[E\left\{Y^1(t)\mathrm{I}(\min(T^1,T^{*0},t)=t)|L=L^*\right\}\right]\\
    &\quad + e^{-\psi_{t-1}}E\left[E\left\{Y^1(t-1)\mathrm{I}(\min(T^1,T^{*0},t)=t-1)|L=L^*\right\}\right]\\
    &\quad - e^{-\psi_{t-1}}E\left[E\left\{Y^1(t-1)\mathrm{I}(\min(T^1,T^{*0},t-1)=t-1)|L=L^*\right\}\right]\\
    &= e^{-\psi_t}E\left[E\left\{Y^1(t)\mathrm{I}(\min(T^1,T^{*0},t)=t)|L=L^*\right\}\right]\\
    &\quad - e^{-\psi_{t-1}}E\left[E\left\{Y^1(t-1)\mathrm{I}(\min(T^1,T^{*0},t)=t)|L=L^*\right\}\right].
\end{align*}
We find that
\begin{align*}
    e^{-\psi_t} &= \frac{\Phi_{0,t} - \Phi_{0,t-1}}{E\left[E\left\{Y^1(t)\mathrm{I}(\min(T^1,T^{*0},t)=t)|L=L^*\right\}\right]}\\
    &\quad + e^{-\psi_{t-1}}\frac{E\left[E\left\{Y^1(t-1)\mathrm{I}(\min(T^1,T^{*0},t)=t)|L=L^*\right\}\right]}{E\left[E\left\{Y^1(t)\mathrm{I}(\min(T^1,T^{*0},t)=t)|L=L^*\right\}\right]}\\
    & = \prod_{j=1}^{t} \left( \frac{E\left[E\left\{Y^1(j-1)\mathrm{I}(\min(T^1,T^{*0},j)=j)|L=L^*\right\}\right]}{E\left[E\left\{Y^1(j)\mathrm{I}(\min(T^1,T^{*0},j)=j)|L=L^*\right\}\right]} \right) \\
    &\quad + \sum_{k=1}^{t} \left[ \frac{\Phi_{0,k} - \Phi_{0,k-1}}{E\left[E\left\{Y^1(k)\mathrm{I}(\min(T^1,T^{*0},k)=k)|L=L^*\right\}\right]}\right.\\
    &\quad\times\left. \prod_{j=k+1}^{t} \left( \frac{E\left[E\left\{Y^1(j-1)\mathrm{I}(\min(T^1,T^{*0},j)=j)|L=L^*\right\}\right]}{E\left[E\left\{Y^1(j)\mathrm{I}(\min(T^1,T^{*0},j)=j)|L=L^*\right\}\right]} \right) \right].
\end{align*}

\subsection{Estimation and inference}
For the additive contrast, we found 
\[
\psi_t = \sum_{s=1}^t\frac{\Phi_s - \Phi_{s-1}}{E\left[P\left\{\min(T^1,T^{*0},s)=s|L=L^*\right\}\right]}.
\]
Estimation of the terms in the numerators can be done as explained in Section \ref{sec: estcond}. For the denominators, we can use the EIF (under $\mathcal{P}_\pi$) to construct an estimator. For an arbitrary $s\in\{1,\ldots,t\}$ and $t\in\{1,\ldots,\tau\}$ the EIF of  $E\left[P\left\{\min(T^1,T^{*0},s)=s|L=L^*\right\}\right] = E\left\{P(T\geq s|A=1,L)P(T\geq s|A=0,L)\right\}$ is 
\begin{align*}
    & P(T\geq s|A=1,L)P(T\geq s|A=0,L) - E\left\{P(T\geq s|A=1,L)P(T\geq s|A=0,L)\right\}\\
    &\qquad + \frac{A}{\pi}\left\{I(T\geq s)-P(T\geq s|A=1,L)\right\}P(T\geq s|A=0,L)\\
    &\qquad + \frac{1-A}{1-\pi}\left\{I(T\geq s)-P(T\geq s|A=0,L)\right\}P(T\geq s|A=1,L).
\end{align*}
Consequently, a one-step estimator is given by the sample average of
\begin{align*}
    & \hat{P}(T\geq s|A=1,L)\hat{P}(T\geq s|A=0,L)+ \frac{A}{\hat{\pi}}\left\{I(T\geq s)-\hat{P}(T\geq s|A=1,L)\right\}\hat{P}(T\geq s|A=0,L)\\
    &\qquad + \frac{1-A}{1-\hat{\pi}}\left\{I(T\geq s)-\hat{P}(T\geq s|A=0,L)\right\}\hat{P}(T\geq s|A=1,L).
\end{align*}
A standard error for the estimator $\hat{\psi}_t$ can be constructed based on its EIF which is given by
\begin{align*}
    \sum_{s=1}^t\frac{\mathrm{EIF}(\Phi_s) - \mathrm{EIF}(\Phi_{s-1})}{\rho_s} - \mathrm{EIF}(\rho_s)\frac{\Phi_s - \Phi_{s-1}}{\rho_s^2},
\end{align*}
where $\rho_s = E\left[P\left\{\min(T^1,T^{*0},s)=s|L=L^*\right\}\right]$.

For the multiplicative contrast, we found 
\begin{align*}
    e^{-\psi_t} &= \frac{\Phi_{0,t} - \Phi_{0,t-1}}{E\left[E\left\{Y^1(t)\mathrm{I}(\min(T^1,T^{*0},t)=t)|L=L^*\right\}\right]}\\
    &\quad + e^{-\psi_{t-1}}\frac{E\left[E\left\{Y^1(t-1)\mathrm{I}(\min(T^1,T^{*0},t)=t)|L=L^*\right\}\right]}{E\left[E\left\{Y^1(t)\mathrm{I}(\min(T^1,T^{*0},t)=t)|L=L^*\right\}\right]}.
\end{align*}
Estimation of $\Phi_{0,t}$ can be done as explained in Section \ref{sec: estcond}. A one-step estimator based on the EIF is obtained by taking the sample average of
\begin{align*}
    &\sum_{s=0}^t \hat{P}(T\geq s|A=1,L)\hat{P}(T\geq s|A=0,L)\hat{E}\left\{Y(s)|A=0,T\geq s,L\right\}\\
    &\qquad + \hat{P}(T\geq s|A=0,L)\frac{A}{\hat{\pi}}\left\{ I(T\geq s) - \hat{P}(T\geq s|A=1,L) \right\}\hat{E}\left\{Y(s)|A=0,T\geq s,L\right\}\\
    &\qquad+\hat{P}(T\geq s|A=1,L)\frac{1-A}{1-\hat{\pi}} \left[Y(s)I(T\geq s)- \hat{P}(T\geq s|A=0,L) \hat{E}\left\{Y(s)|A=0,T\geq s,L\right\}\right]\\
    &-I(t>s)\Big(  \hat{P}(T> s|A=1,L)\hat{P}(T> s|A=0,L)\hat{E}\left\{Y(s)|A=0,T> s,L\right\}\\
    &\qquad + \hat{P}(T> s|A=0,L)\frac{A}{\hat{\pi}}\left\{ I(T> s) - \hat{P}(T> s|A=1,L) \right\}\hat{E}\left\{Y(s)|A=0,T> s,L\right\}\\
    &\qquad+\hat{P}(T> s|A=1,L)\frac{1-A}{1-\hat{\pi}} \left[Y(s)I(T> s)- \hat{P}(T> s|A=0,L) \hat{E}\left\{Y(s)|A=0,T> s,L\right\}\right]   \Big).
\end{align*}
The denominators $E\left[E\left\{Y^1(t)\mathrm{I}(\min(T^1,T^{*0},t)=t)|L=L^*\right\}\right]$ are equal to $\theta_{t,t}$ and the numerator in the second term is $\theta_{t-1,t}$, with notation from Appendix B where also the corresponding EIFs are determined. A one-step estimator for $\theta_{s,u}$ is given by the sample average of
\begin{align*}
    &\hat{P}(T> u|A=0,L)\hat{P}(T> u|A=1,L)E\{Y(s)|A=1,L,T>u\}\\
    & + \frac{1-A}{1-\hat{\pi}}\left\{I(T>s) - \hat{P}(T>s|A=0,L) \right\}\hat{P}(T> u|A=1,L)\hat{E}\{Y(s)|A=1,L,T>u\}\\
    &+ \hat{P}(T>s|A=0,L)\frac{A}{\hat{\pi}}\left[Y(s)I(T>s)-\hat{P}(T> u|A=1,L)\hat{E}\{Y(s)|A=1,L,T>u\}\right].
\end{align*}
Now, $e^{-\psi_t}$ can be recursively estimated based on the estimate for $e^{-\psi_{t-1}}$ (and $e^{-\psi_0}=1$) and using one-step estimators for $\Phi_{0,t}$ and for $\theta_{s,u}$. A standard error can be obtained as $n^{-1/2}$ times the sample standard deviation of $\mathrm{EIF}(e^{-\psi_t})$ which is given by
\begin{align*}
    &\frac{\mathrm{EIF}(\Phi_{0,t}) - \mathrm{EIF}(\Phi_{0,t-1})}{\theta_{t,t}} - \mathrm{EIF}(\theta_{t,t})\frac{\Phi_{0,t} - \Phi_{0,t-1}}{\theta_{t,t}^2}\\
    &\quad + \mathrm{EIF}(e^{-\psi_{t-1}})\frac{\theta_{t-1,t}}{\theta_{t,t}}+ e^{-\psi_{t-1}}\left\{\frac{\mathrm{EIF}(\theta_{t-1,t})}{\theta_{t,t}}+\mathrm{EIF}(\theta_{t,t})\frac{\theta_{t-1,t}}{\theta_{t,t}^2}\right\}.
\end{align*}

\newpage
\section*{Appendix G: Data analysis - additional information}
\begin{figure}[h]
    \centering
    \includegraphics[width=0.8\linewidth]{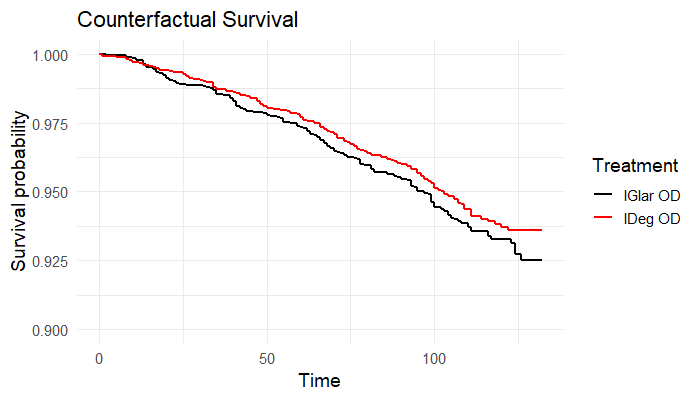}
    \caption{\textbf{Counterfactual Survival.} 
    This figure displays the estimated counterfactual survival probabilities for insulin degludec (IDeg OD) and insulin glargine (IGlar OD) over a period of 132 weeks.}
    \label{fig:devote_survival}
\end{figure}

To contextualize the distribution used for weighting the CPLOT estimands, we performed a counterfactual survival analysis where ``survival'' specifically represents the time until death, with dropout and the administrative end of study treated as censoring events. The resulting curves represent the counterfactual survival probability in the study population. As shown in Figure \ref{fig:devote_survival}, both treatment arms maintain high retention rates, with survival probabilities exceeding $92.5\%$. Results were obtained using the \texttt{CFsurvival} package in R, which provides non-parametric estimation and inference for counterfactual survival functions under a specified treatment strategy \citep{westling2024inference}.

\end{document}